\documentclass[letterpaper,twocolumn,10pt]{article}
\usepackage{usenix2019_v3}

\usepackage{paralist}
\usepackage{color,soul}
\usepackage{hyperref}
\usepackage{url}

\usepackage{times}
\usepackage{graphicx}
\usepackage{rotating}
\usepackage{fancybox}
\usepackage{multirow}
\usepackage{listings}
\usepackage{booktabs}
\usepackage{array}
\usepackage{comment}
\usepackage{alltt}
\usepackage{subcaption}
\usepackage[usestackEOL]{stackengine}
\usepackage{colortbl}
\usepackage{titlesec}
\usepackage{fancyhdr}
\usepackage{xspace}
\usepackage{amssymb}
\usepackage{pifont}
\usepackage{csquotes}

\newcommand*\rot[1]{\hbox to1em{\hss\rotatebox[origin=br]{-30}{#1}}}

\titlespacing*\section{0pt}{5pt plus 2pt minus 2pt}{3pt plus 1pt minus 1pt}
\titlespacing*\subsection{0pt}{4pt plus 2pt minus 1pt}{3pt plus 1pt minus 1pt}
\titlespacing*\subsubsection{0pt}{2pt plus 2pt minus 1pt}{1pt plus 1pt minus 1pt}

\titlespacing{\paragraph}{%
  0pt}{%
  0.25\baselineskip}{%
  1em}%

\date{}

\newcommand{\scheme}{\textsc{Membuster}\xspace}
\newcommand{\pinning}{critical page whitelisting}
\newcommand{\etal}{{\it et al.}}
\newcommand{\cpumodel}{{i5-8400}}
\newcommand{\gb}{{GB}}
\newcommand{\gib}{{GiB}}
\newcommand{\mb}{{MB}}
\newcommand{\kb}{{KB}}

\newcommand{\squeezing}{cache squeezing\xspace}
\newcommand{\Squeezing}{Cache squeezing\xspace}
\newcommand{\priming}{cross-core priming\xspace}
\newcommand{\hunspell}{Hunspell\xspace}
\newcommand{\memcached}{Memcached\xspace}
\newcommand{\cmark}{\ding{51}\xspace}%
\newcommand{\xmark}{\ding{55}\xspace}%

\newcommand{\primeprobe}{\textsc{Prime+Probe}\xspace}
\newcommand{\flushreload}{\textsc{Flush+Reload}\xspace}
\newcommand{\flushflush}{\textsc{Flush+Flush}\xspace}

\def \debug{a}

\ifx \debug \undefined

\newcommand{\fixme}[1]{{}}
\newcommand{\raluca}[1]{{}}
\else

\newcommand{\fixme}[1]{\hl{ #1}}
\newcommand{\raluca}[1]{{\bf \textcolor{red}{Raluca:  #1}}}
\fi

\begin{document}

\title{An Off-Chip Attack on Hardware Enclaves via the Memory Bus}

\author{
{\rm Dayeol Lee}\\
UC Berkeley
\and 
{\rm Dongha Jung} \\
SK Hynix
\and
{\rm Ian T. Fang}\\
UC Berkeley
\and 
{\rm Chia-Che Tsai} \\
Texas A\&M University
\and
{\rm Raluca Ada Popa} \\
UC Berkeley
}%

\maketitle

\thispagestyle{fancy}

\begin{abstract}

This paper shows how an attacker can break the confidentiality of a hardware enclave with \scheme,
an {\em off-chip} attack based on snooping the memory bus. 
An attacker with physical access can observe an unencrypted address bus 
and extract fine-grained memory access patterns of the victim.
\scheme is qualitatively different from prior on-chip attacks to enclaves
and is more difficult to thwart. 

We highlight several challenges for \scheme.
First, DRAM requests are only visible on the memory bus at
last-level cache misses.
Second, the attack needs to incur minimal interference or overhead to the victim
to prevent the detection of the attack.
Lastly, the attacker needs to reverse-engineer
the translation between virtual,
physical, and DRAM addresses to perform a robust attack.
We introduce three techniques, \textit{\pinning}, \textit{\squeezing}, and \textit{oracle-based fuzzy matching algorithm} 
to increase cache misses for memory accesses that are useful for the attack, with no detectable interference to the victim, and to convert memory accesses to sensitive data. 
We demonstrate \scheme on an Intel SGX CPU to leak
confidential data from two applications: \hunspell and \memcached.
We show that a single uninterrupted run of the victim can leak most of the sensitive data with high accuracy.

\end{abstract}

\section{Introduction}
\label{sec:intro}

Hardware enclaves~\cite{sgx, mckeen13sgx, lie2003implementing, keystone, sanctum}
provide secure execution environments to protect sensitive code and data.
A hardware enclave has a small trusted computing base (TCB)
including the trusted hardware and program
and assumes a strong threat model where even a privileged attacker (e.g., hypervisor, OS)
cannot break the confidentiality and integrity of the execution.
In such a threat model, 
the attacker cannot physically attack the internals of the processor package, but can attempt to tamper with or observe the externals of the processor (e.g., Cold-Boot attacks~\cite{cold-boot}). 
As a result, hardware enclaves are attractive for protecting privacy-sensitive workloads such as
database~\cite{enclavedb}, big data
~\cite{vc3, m2r, securekeeper}, blockchains~\cite{teechain, proof-of-luck, towncrier, ekiden, tesseract}, and machine
learning~\cite{sgx-ml, privado}.

\begin{figure}
    \centering
    \includegraphics[width=0.9\linewidth]{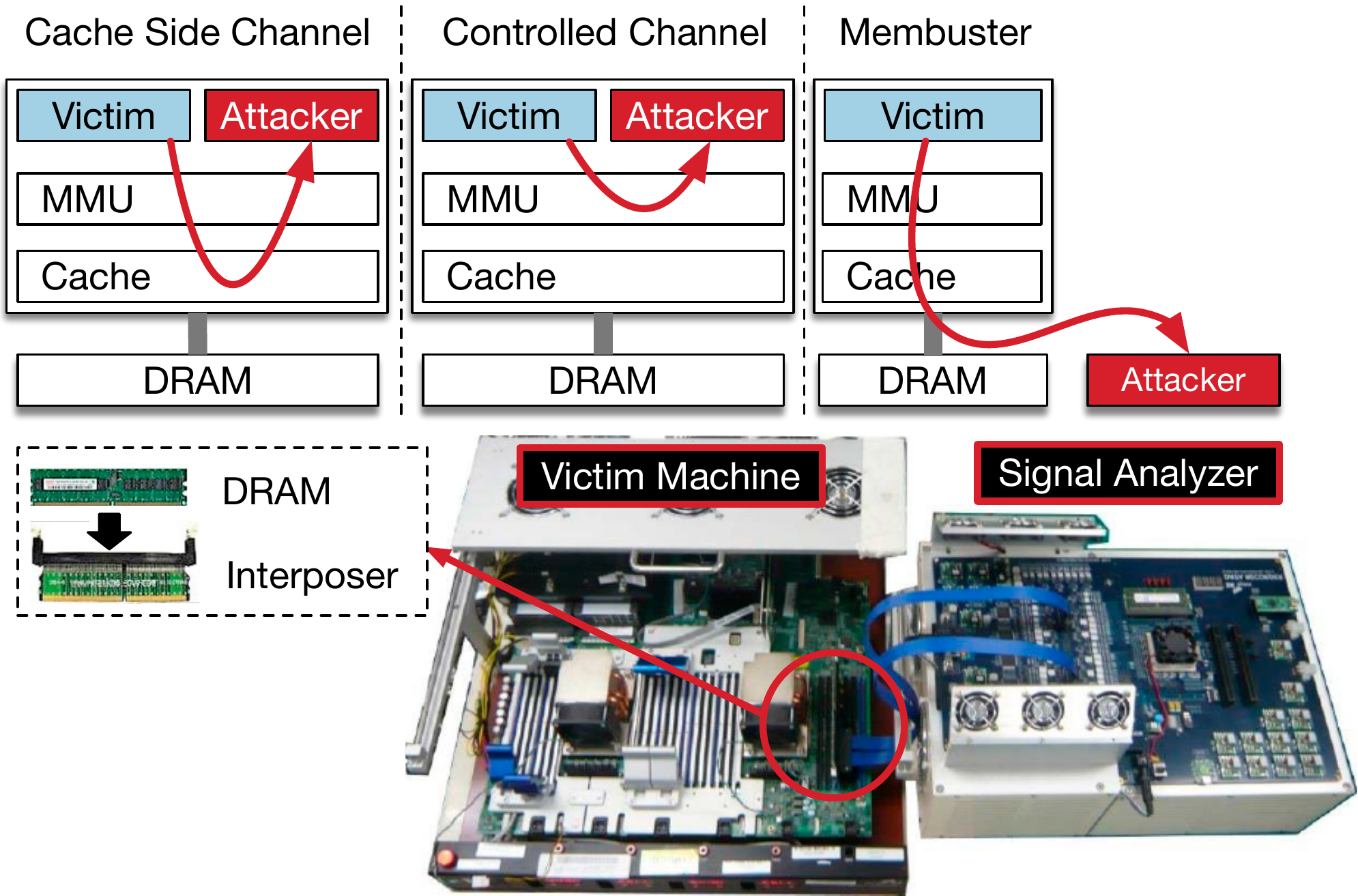}
    \caption{On-chip side channels compared to \scheme. The cache side-channel attack leaks addresses through a shared cache, whereas the controlled-channel attack uses adversarial memory management. \scheme leaks addresses directly through the off-chip memory bus.
    The photo shows an example hardware setup for the attack.}
    \label{fig:offchip_channel}
\end{figure}

Along with the proliferation of hardware enclaves, many side-channel attacks against them
have been discovered~\cite{foreshadow, wang17side-channel, brasser17cache-attack, conceal, cachezoom, VanBulck2017attack}.
Side-channel attacks leak sensitive information from enclaves via architectural or microarchitectural states.
For instance, controlled-channel attacks~\cite{controlled} use the OS privilege to trigger page faults for memory access on different pages, to
reconstruct secrets from page-granularity access patterns inside the victim program. 
We categorize these attacks as \textit{on-chip side-channel attacks}, 
where the attacker uses adversarial or shared on-chip components to reveal memory addresses accessed by the victim
(Figure~\ref{fig:offchip_channel}).

An attacker who can {\em physically} access the machine can perform an
\textit{off-chip side-channel attack} that directly observes the memory addresses on the
\textit{memory bus}.
The memory bus, which consists of a \textit{data bus} and an \textit{address bus}, delivers
memory requests from a CPU to an off-chip DRAM.
Although the CPU encrypts the data of an enclave, all the addresses still leave the CPU unencrypted,
allowing the attacker to infer program secrets from the access patterns.
Since off-the-shelf DRAM interfaces do not support address bus encryption,
no existing hardware enclave can prevent physical attackers from observing the memory address bus.

Several studies have hinted at the possibility of attacks based on the memory address bus~\cite{phantom,sgx-explained,xbox}.
Costan~\etal~\cite{sgx-explained} suggest the possibility of tapping the address bus, 
but acknowledge that they are not aware of any successful example of the attack.
Maas~\etal~\cite{phantom} suggest that an attacker who can collect physical memory traces of a database server can distinguish two different SQL queries operating on the same dataset.
However, to the best of our knowledge, no work has shown how such a side channel can be exploited to break the confidentiality of an enclave.

In this paper, we present \scheme, an off-chip side-channel attack on the memory address bus.
We show that \scheme can be a substantial threat
to hardware enclaves because of its unique traits compared to the existing on-chip attacks (\S\ref{s:comparison}).
The need for off-chip access, despite being a disadvantage, advantages the attacker as it makes 
\scheme much harder to mitigate with \textit{protected-access} solutions (Table~\ref{tab:compare}).
Recently, a wide range of tools~\cite{Varys,Hyperrace,Cloak,tsgx,chen17side-channel} have been developed for mitigating on-chip side-channel attacks for enclaves with a reasonable overhead. 
These tools either partition the resources (e.g., cache) to prevent an attacker from learning information
via shared resources or intercept actions (e.g., page faults) to
prevent an attacker from observing the side channels. At their core, these solutions attempt to protect the memory accesses from an attacker's sight. 

However, these protected-access solutions do not prevent \scheme,
which observes the memory addresses off-chip
and thus can bypass the protection
of any on-chip solutions.
To prevent \scheme on the current hardware enclave design, one must
{\em hide} the accessed memory addresses, by making the enclave execution {\em oblivious}
to the secret data.  
This requires either using oblivious algorithms~\cite{WangNLCSSH14} inside the enclave or running the enclave atop an ORAM~\cite{pathoram,  Oblix}. 
Both mechanisms bring
significant performance overhead to the enclave. 
An alternative is to change the CPU and DRAM design to encrypt the address bus, but implementing a decryption module in DRAM can be expensive~\cite{InvisiMem, ObfusMem}.

We describe the challenges to perform a robust off-chip attack
as follows:
{\bf (1) Address Translation.}
The attacker needs to translate the DRAM requests into the physical addresses
by reverse-engineering the mapping and to further translate them
into virtual addresses of the victim enclave;
{\bf (2) Lossy Channel.}
The attacker only sees DRAM requests when cache misses or write-back occurs. Since most modern CPUs have a large last-level cache (LLC), a significant portion of memory accesses do not issue any DRAM requests.
We show why simple methods such as \textit{priming} the cache
does not incur sufficient cache misses needed for the attack;
{\bf (3) Unusual Behaviors in SGX. }
SGX has unique memory behaviors
which increase the difficulty of the attack.
For example, we show that common architectural features such as disabling the cache
do not work in SGX.
We also find that \textit{paging} in SGX hides most of the memory accesses.

We first show how an attacker can translate the DRAM requests, 
and can filter out irrelevant addresses to
leave only the \textit{critical} addresses that are useful for the attack.
Then, we introduce two techniques, {\em \pinning} (\S\ref{sec:improvements:pinning}) and {\em \squeezing} (\S\ref{s:squeezing}), to increase useful cache misses by
thwarting page swaps and
shrinking the effective cache for the critical addresses. With more cache misses, the attacker can observe more DRAM requests. 
These techniques do not cause detectible interference to the victim,
and can be combined with cache priming to make more memory accesses visible
to the attacker.
Our {\em oracle-based fuzzy matching algorithm} (\S\ref{sec:matching}) can create an ``oracle'' of the secret-to-access-pattern mapping,
to identify the sensitive accesses from a sizable memory bus trace.
We then extract the sensitive data from the noisy memory accesses by fuzzy-matching the accesses against the oracle.
We further show that {\em hardware prefetching} can increase the efficiency of this algorithm in \scheme{}.

We demonstrate the attack by attaching Dual In-line Memory Module (DIMM) interposer
to a production system with an SGX-enabled Intel processor and a commodity DDR4 DRAM.
We capture the memory bus signals to perform an off-line analysis. 
We use two applications, Hunspell and Memcached, to demonstrate the attack.
Finally, we show the scalability of our techniques by simulating the attack in modified QEMU~\cite{qemu}.

To summarize, the contributions of this paper are as follows:
\begin{compactitem}

    \item The setup of an off-chip side-channel attack on hardware enclaves and identification of the challenges for launching the attack robustly.    
    
    \item Effective techniques for maximizing the side-channel information with no detectible interference nor order-of-magnitude performance overhead to the victim program.
    
    \item A fuzzy comparison algorithm for converting the address trace
    collected on the memory bus
    to program secrets. 

    \item Demonstration and experimentation of the attack on an actual Intel SGX CPU. To our best knowledge, it is the first work that shows the practicality of the attack.

\end{compactitem}

The security implications of the off-chip side-channel attacks can be pervasive because such a channel exists on almost every secure processor with untrusted memory. We hope to motivate  further research by alarming the community about the practicality and severity of such  attacks.

\section{Background and Related Work}
\label{sec:background}

In this section, we discuss the background, including hardware enclaves, known on-chip side-channel attacks on SGX, and the related defenses.

\subsection{Intel SGX}

We choose Intel SGX~\cite{sgx-manual} as the primary attack target because Intel SGX has the most mature implementation and the strongest threat model against untrusted DRAM. 
SGX is a set of instructions for supporting hardware enclaves introduced in the Intel 6th generation processors.
SGX assumes that only the processor package is trusted;
all the off-chip hardware devices, including the DRAM and peripheral devices, are considered potentially vulnerable or compromised.
The threat model of SGX also includes physical attacks
such as Cold-Boot Attacks~\cite{cold-boot}, which can observe sensitive data from residuals inside DRAM.

An Intel CPU with SGX contains a memory encryption engine (MEE), which 
encrypts and authenticates the data stored in a dedicated physical memory range called the {\it enclave page cache} (EPC).
The MEE encrypts data blocks and generates authentication tags
when sending the data outside the CPU package to be stored inside the DRAM.
To prevent roll-back attacks, the MEE also stores a version tree of the protected data blocks,
with the top level of the tree stored inside the CPU.
For Intel SGX, EPC is a limited resource; the largest EPC size currently available on an existing Intel CPU is 93.5 MB, out of 128 MB Processor's Reserved Memory (PRM).
The physical pages in EPC, or EPC pages,
are mapped to virtual pages
in enclave linear address ranges (ELRANGEs) by the untrusted OS.
If all concurrent enclaves require more virtual memory than the EPC size, 
the OS needs to swap the encrypted EPC pages into regular pages. 

However, even with MEE, Intel SGX does not encrypt the addresses on the memory bus. 
As previously discussed, changing the CPU to encrypt the addresses 
requires implementing the encryption logic on DRAM, and thus requires new technologies such as Hybrid Memory Cube (HMC)~\cite{InvisiMem, ObfusMem}.

The unencrypted address bus opens up a universal threat to
hardware enclaves with external encrypted memory.
Komodo~\cite{komodo}, ARM CryptoIsland~\cite{cryptoisland}, Sanctum~\cite{sanctum}, and Keystone~\cite{keystone} do not
encrypt data for an external memory by default.
AMD SEV~\cite{amd-sev} allows hypervisor-level memory encryption, but also does not encrypt addresses.

\definecolor{Gray}{gray}{0.90}
\definecolor{LB}{HTML}{92c5de}
\definecolor{LR}{HTML}{ffffff}

\begin{table}
\footnotesize
\centering
\begin{tabular}{lllllll}
\hline
\textbf{} & \textbf{\rot{Brasser et al.~\cite{brasser17cache-attack}}} & \textbf{\rot{Schwarz et al.~\cite{conceal}}} & \textbf{\rot{CacheZoom~\cite{cachezoom}}} & \textbf{\rot{\textsc{Flush}-based~\cite{VanBulck2017attack}}} & \textbf{\rot{Controlled~\cite{controlled}}} & \textbf{\rot{\scheme}} \\ \hline
\bf Software-Only       & 
	\cellcolor{LB} \cmark & 
	\cellcolor{LB} \cmark &
	\cellcolor{LB} \cmark &
	\cellcolor{LB} \cmark & 
	\cellcolor{LB} \cmark & 
	\cellcolor{LR} \xmark 
	\\ \hline
\bf Protected-Access Fix~\cite{Varys,Hyperrace,Cloak,tsgx,chen17side-channel}  & 
	\cellcolor{LR} \cmark &
	\cellcolor{LR} \cmark &
	\cellcolor{LR} \cmark &
	\cellcolor{LR} \cmark &
	\cellcolor{LR} \cmark &
	\cellcolor{LB} \xmark 
	\\ \hline
Root Adversary       &
	\cellcolor{LR} \cmark & 
	\cellcolor{LB} \xmark &
	\cellcolor{LR} \cmark &
	\cellcolor{LR} \cmark & 
	\cellcolor{LR} \cmark & 
	\cellcolor{LR} \cmark 
	\\ \hline
Noiseless           &
	\cellcolor{LR} \xmark & 
	\cellcolor{LR} \xmark &
	\cellcolor{LR} \xmark &
	\cellcolor{LB} \cmark & 
	\cellcolor{LB} \cmark & 
	\cellcolor{LB} \cmark 
	\\ \hline
Lossless            &
	\cellcolor{LR} \xmark &
	\cellcolor{LR} \xmark &
	\cellcolor{LR} \xmark &
	\cellcolor{LB} \cmark & 
	\cellcolor{LB} \cmark & 
	\cellcolor{LR} \xmark 
	\\ \hline
Fine-Grained (64B vs. 4KB)   &
	\cellcolor{LB} \cmark &
	\cellcolor{LB} \cmark &
	\cellcolor{LB} \cmark &
	\cellcolor{LR} \xmark &
	\cellcolor{LR} \xmark & 
	\cellcolor{LB} \cmark 
	\\ \hline
No Interference (e.g., AEX)  &
	\cellcolor{LB} \cmark  &
	\cellcolor{LB} \cmark &
	\cellcolor{LR} \xmark & 
	\cellcolor{LR} \xmark & 
	\cellcolor{LR} \xmark & 
	\cellcolor{LB} \cmark 
	\\ \hline
Low Overhead     & 
	\cellcolor{LB} \cmark  & 
	\cellcolor{LB} \cmark &
	\cellcolor{LR} \xmark &
	\cellcolor{LR} \xmark & 
	\cellcolor{LR} \xmark & 
	\cellcolor{LB} \cmark 
	\\ \hline

\end{tabular}
\caption{This work (\scheme) compared to previous side-channel attacks on SGX. The two boldface rows illustrate what we perceive to be the most important distinctions. The colored cell indicates the attacker has the advantage.}    
    \label{tab:compare}
\end{table}

\subsection{Comparison with Existing Attacks}
\label{s:comparison}

In this section, we discuss how \scheme can be a substantial threat
to hardware enclaves because of its unique traits.
We compare \scheme{} with various on-chip side-channel attacks on SGX~\cite{brasser17cache-attack,cachezoom,conceal,VanBulck2017attack,controlled} in Table~\ref{tab:compare}.

\subsubsection{Side Channel Attacks on SGX}

\noindent {\textbf{\primeprobe.}} 
A shared cache hierarchy allows an adversary to infer memory access patterns of the victim using known techniques such as \primeprobe~\cite{Osvik2006,liu15cache-attack}.
However, in \primeprobe{},
the attacker usually
cannot reliably distinguish the victim's accesses from \textit{noises} of other processes.
The \primeprobe channels are also \textit{lossy}, as the attacker may miss some of victim's accesses while probing. 

Brasser \etal~\cite{brasser17cache-attack} demonstrate \primeprobe on Intel SGX 
without interfering with the enclave, but the attack requires running the victim program repeatedly 
to compensate for its noise and signal loss.
Schwarz \etal~\cite{conceal} show that the attacker can alleviate the noise by identifying cache sets
that are critical to the attack.
This technique can be applied to applications that have data-dependent accesses in a small number of cache sets.
CacheZoom~\cite{cachezoom} also uses \primeprobe but minimizes the noise by inducing Asynchronous Exits (AEXs) every few memory accesses in the victim.
This incurs a significant overhead on enclaves, and also makes the attack easily detectable
\cite{chen17side-channel}.

\noindent \textbf{Flush-based Side Channels.}
Other techniques such as \flushreload~\cite{flush-reload} and \flushflush~\cite{flush-flush}
use a shared cache block between the attacker and the victim 
to create a noiseless and lossless side channel.
However, these techniques cannot be directly applied to enclave memory, 
because an enclave does not share the memory with other processes.
However, these techniques can still be used to
observe the page table walk for enclave addresses~\cite{VanBulck2017attack}.
Specifically, the attacker can monitor the target page tables with a tight \flushreload loop.
As soon as the loop detects page table activities, the attacker interrupts the victim
and infers page-granularity addresses.
Similar to CacheZoom, this attack incurs a significant AEX overhead and thus can be detected by the victim.

\noindent \textbf{Controlled Channels.}
Controlled-channel attacks~\cite{controlled} take advantage of the adversarial memory management of the untrusted OS, 
to capture the access patterns of an SGX-protected execution.
Even though Intel SGX masks the lower 12 bits of the page fault addresses to the untrusted OS, controlled-channel attacks use sequences of virtual page numbers to differentiate memory accesses within the same page.
The controlled channel is noiseless and lossless but can be detected and mitigated as it incurs a page fault for each sequence of accesses on the same page~\cite{tsgx,Varys}.

\subsubsection{Advantages of \scheme}

As shown in Table~\ref{tab:compare}, \scheme creates a noiseless side channel by filtering out all of the non-victim memory accesses, leaving only addresses that are useful for the attack.
It can observe memory accesses with cache line granularity.
Also, \scheme does not incur interference such as AEX or page fault to the victim and 
needs not to incur an order-of-magnitude overhead.

Several recent mechanisms, such as Varys~\cite{Varys}, Hyperrace~\cite{Hyperrace}, Cloak~\cite{Cloak}, T-SGX~\cite{tsgx}, or D\'ej\`a Vu~\cite{chen17side-channel}, have been proposed to prevent
the attacker from observing memory access patterns in the victim.
In general, \primeprobe can be mitigated by partitioning the cache to shield the victim from on-chip attackers.
This does not defeat an off-chip attacker who directly observes DRAM requests.
T-SGX~\cite{tsgx} and D\'ej\`a Vu~\cite{chen17side-channel} have proposed to use the Intel Transactional Synchronization Extensions (TSX)
to prevent AEX or page faults from an enclave.
These techniques are based on thwarting the interference (e.g., AEX, page faults) that causes
the side channels~\cite{cachezoom,VanBulck2017attack,controlled}.
However, \scheme does not
incur such interference on enclaves, and thus cannot be thwarted through similar approaches.
To our best knowledge, there is no reliable way to detect or mitigate \scheme using existing on-chip measures.

\subsubsection{Related Work}

\noindent \paragraph{Other On-Chip Attacks.}
Other on-chip attacks worth mentioning are speculative-based execution side channels like Foreshadow~\cite{foreshadow} or ZombieLoad~\cite{SGX:attack:ZombieLoad}, branch shadowing side channels~\cite{sgxattacks-lee:branches}, denial-of-service attacks (e.g.,  Rowhammer~\cite{SGXattack:rowhammer:SGXbomb:DOS:Jang,drammer}), or rollback attacks~\cite{SGX:LCM:defense:rollback:Brandenburger:2018, SGX:ROTE:defense:rollback:Matetic:2017}. 

\noindent \paragraph{Other Off-Chip Side-Channel Attack.}

DRAM row buffers can be exploited as side-channels between cores or CPUs, as demonstrated in DRAMA~\cite{drama}. DRAMA shows that by observing the latency of reading or writing to DRAM, the attacker can infer whether the victim has recently accessed the data stored in the same row. 
DRAMA shows how a software-only attacker can use DRAM row buffers as covert channels or side channels.
\scheme further explores how the attacker can directly use the address bus as a side channel.

\section{\scheme{}}
\label{sec:attack}

In this section, we describe the basic attack model of \scheme{}. In further sections, we will refine and improve the attack.
At a high level, the attacker first sets up an environment to collect the DRAM signals and waits until the victim executes some code containing data-dependent memory accesses.
The attacker translates the collected signals into cache-line granularity virtual addresses.

\subsection{Threat Model}
\label{sec:threat}

We assume the standard Intel SGX threat model in which nothing but the CPU package and the victim program is trusted.
Everything else, including the OS or other applications, is untrusted and can be controlled by the attacker.
External hardware devices are also untrusted, so the attacker can tap the address bus to the external DRAM.
For the advanced techniques discussed in \S\ref{sec:improvement},
the attacker may also use the root privilege to install the modified SGX driver.

To tap the memory bus, the attacker needs to have physical access to the machine where the victim is running.
Such an assumption eliminates the possibility of remote attacks through either cloud environments or network connections.
The candidates who may perform \scheme could be two types. On the server-side, these may include the employees of a cloud provider, or IT administrators of an institution, who act as insiders to leak sensitive information. 
On the client-side, end users may want to attack the local hardware enclaves, which protect proprietary data (e.g., licenses,  digital properties, etc).
We assume that the attacker has enough budget and knowledge to acquire and install the DIMM interposer for the attack described in \S\ref{sec:hardware-setup}.
This might be an obstacle for the general public, but we claim that the cost is manageable if the attacker has a strong motivation for obtaining the data.

Like in the controlled channel and cache side channels, \scheme{}  assumes that the adversary has knowledge of the victim application, by either consulting the source code or reverse-engineering the application.
The adversary is also aware of the runtime used by the victim application for platform support,
such as the SDK libraries, library OSes, or shield systems.
In our experiments, we use Graphene-SGX~\cite{graphene} for platform support of the victim applications.
Address Space Layout Randomization (ASLR) in the library OSes or the runtimes may complicate the extraction of secret information
but generally is insufficient to conceal the access patterns completely~\cite{controlled}.
ASLR offered by the host kernel is irrelevant because a hostile host kernel can either control or monitor the addresses where the victim enclaves are loaded.

\subsection{Hardware Setup for the Attack}
\label{sec:hardware-setup}

\begin{figure}
    \centering
    \includegraphics[width=0.9\linewidth]{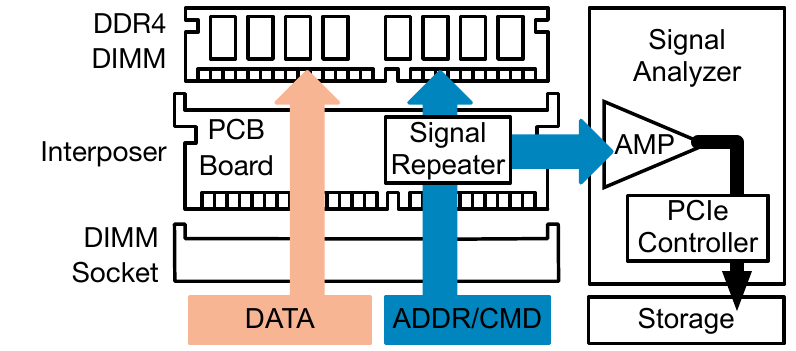}
    \caption{Hardware setup for a memory bus side-channel attack.
    DIMM interposer collects the bus signals and sends them to the signal analyzer.
    The attacker can use the analyzed signals to learn the memory access pattern of the victim.}
    \label{fig:attack-setting}
\end{figure}

Figure \ref{fig:attack-setting} shows a detailed hardware setup for the \scheme{} attack. 
The hardware setup may vary on different CPU models and vendors.
The attacker installs an {\em interposer} on the DIMM socket prior to system boot. 
The interposer is a custom printed circuit board (PCB) that can be placed between the DRAM and the socket.
The interposer contains a signal repeater chip which duplicates the command bus signals and sends them to a {\em signal analyzer}.
The analyzer amplifies the signals and then outputs the signals to a storage server through a PCIe interface.

In the rest of the section, we will highlight the key requirements in successfully performing the attack.

\noindent{\bf Sampling Rate.}
The sampling rate of the interposer needs to be equal or higher than the clock rate of the DIMM in order to capture all the memory requests.
A standard DDR4 clock rate ranges from 800 to 1600 MHz, while a DIMM typically supports between 1066 (DDR4-2133) and 1333 (DDR4-2666) MHz.
To match with the sampling rate,
the attacker can lower the DIMM clock rate if it is configurable in the BIOS.

\noindent{\bf Recording Bandwidth.} The sampling rate also determines the \textit{recording bandwidth}.
For example, DDR4-2400 (1200 MHz) has a 32-bit address and a 64-bit data bus,
thus the recording bandwidth for the address bus is $1200$ Mbps$\times 32$ bits $= 4.47$ \gib/s.
For reference, the data bus of a DDR has a $2\times$ transfer rate, as well as a $2\times$ transfer size. Hence, the bandwidth for logging all the data on DDR4-2400 will be 17.88 \gib/s.

\noindent{\bf Acquisition Time Window. }
The \textit{acquisition time window} (i.e., the maximum duration for collecting the memory commands) determines the maximum length of execution that the attacker can observe.
The acquisition time window equals the \textit{acquisition depth} (i.e., the analyzer's maximum capacity of processing a series of contiguous sample) divided by the recording bandwidth of the interposer.
For example, with $64$ \gib{} acquisition depth, the analyzer can process and log the commands from DDR4-2400 up to $\sim 14$ seconds.

We surveyed several vendors which offer DIMM analyzers~\cite{jki-analyzer,protocol-analyzer,protocol-analyzer2} for purchase or rental.
Among them, the maximum sampling rate can reach $1200$-$1600$ MHz, and the acquisition depth typically ranges between $4$-$60$ \gib.
One of the devices~\cite{jki-analyzer} can extend the acquisition time window to $> 1$ hour by attaching 16 TB SSD and streaming the compressed log via PCIe at 4.8 \gib/s.
Another device~\cite{protocol-analyzer2} does not disclose the memory depth but specifies that it can capture up to 1G ($10^9$) samples.
The cost of the analyzer varies depending on the sampling rate and the acquisition depth.
At the time of writing, \emph{Kibra 480}~\cite{protocol-analyzer} (1200 MHz, $4$ \gib) costs \$6,500 per month, 
\emph{MA4100}~\cite{protocol-analyzer2} (1600 MHz, $1$G-samples) costs \$8,000 per month,
and \emph{JLA320A}~\cite{jki-analyzer} (1600 MHz, $64$ \gib) costs \$170,000 for purchase.

\subsection{Interpreting DRAM Commands}

Once the attacker has finished setting up the environment, she can collect the DRAM signals at any point in time, and analyze the trace off-line.
As the first step, the attacker interprets the DRAM commands collected from the interposer.

A modern DRAM contains multiple banks that are separated into bank groups. 
Within each bank, data (often of the same size as the cache lines) are located by rows and columns. 
Each bank has a row buffer (i.e., a sense amplifier) for temporarily holding the data of a specific row when the CPU needs to read or write in the row. 
Because only one row can be accessed in a bank at a time, the CPU needs to reload the row buffer when accessing a data block in another row.

The log collected from the DRAM interposer typically consists of the following commands: 

\begin{compactitem}
    \item {\tt ACTIVATE(Rank,Bank,BankGroup,Row)}:
    Activating a specific row in the row buffer for a certain rank, bank, and bank group.
    \item {\tt PRECHARGE(Rank,Bank,BackGroup)}:
    Precharging and deactivating the row buffer for a certain rank, bank, and bank group.
    \item {\tt READ(Rank,Bank,BankGroup,Col)}:
    Reading a data block at a specific column in the row buffer.
    \item {\tt WRITE(Rank,Bank,BankGroup,Col)}:
    Writing a data block at a specific column in the row buffer.
\end{compactitem}

Other commands such as PDX (Power Down Start), PDE (Power Down End), and AUTO (Auto-recharge) are irrelevant to the attack and thus omitted from the logs.

Based on the DRAM commands, we can construct the rank, bank, row, and column of each trace,
by simply tracing the activated row within each bank.
Note that the final traces are also time-stamped by the clock counter of the analyzer.
The result of the translation is a sequence of logs containing the timestamp, access type (read or write), rank, bank, row, and column in the DRAM.

\subsection{Reverse-engineering DRAM Addressing}
\label{sec:attack:pa-to-dram}

A physical address in the CPU does not linearly map to a DRAM address consisting of rank, bank, row, and column.
Instead, the memory controller translates the address to maximize DRAM bank utilization and minimize the latency.
The translation logic heavily depends on the CPU and DRAM models, and Intel does not disclose any information.
Thus, the attacker needs to reverse-engineer the internal translation rule for the specific set of hardware.
This has been also done by a previous study~\cite{drama}.

We use the traces collected from the DRAM interposer to reverse-engineer the addressing algorithm of an Intel CPU.
For attacking the enclaves, we only need a part of the addressing algorithm that affects the range of the enclave page cache (EPC). We write a program running inside an enclave, which probes the DRAM addresses translated from the EPC addresses.
The probing program allocates a heap space larger than the EPC size (93.5MB). For every cache line in the range, the program generates cache misses by repeatedly flushing the cache line and fetching it into the cache. By accessing each cache line multiple times, we can differentiate the traces caused by probing
from other memory accesses in the background
and minimize the effect of re-ordering by the CPU's memory controller.
The techniques in \S\ref{sec:attack:pa-to-va} are also needed for translating the probed virtual addresses to physical addresses.

\begin{figure}
    \centering
    \includegraphics[width=0.6\linewidth]{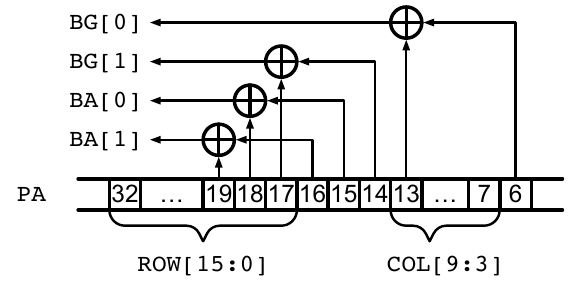}
    \caption{The reverse engineered addressing function of the \cpumodel{} CPU. The function translate a physical address ({\tt PA}) to the Bank Group ({\tt BG}), Bank Address ({\tt BA}), Row ({\tt ROW}) and Column ({\tt COL}) within the DRAM.
    }%
\label{fig:pa-to-dram}
\end{figure}

Using the DRAM traces generated by probing cache lines inside the EPC, we can create a direct mapping between the physical addresses and DRAM addresses (ranks, banks, bank groups, rows, and columns). We further deduce the addressing function of the target CPU (\cpumodel{}), by observing the changing bits in the physical addresses when DRAM addresses change.
We conclude that the addressing function on \cpumodel{} is as shown in Figure~\ref{fig:pa-to-dram}. Other CPU models may implement a different addressing function, and reverse-engineering should be done for each CPU model beforehand.

\subsection{Translating PA to VA}
\label{sec:attack:pa-to-va}

In order to extract the actual memory access pattern of the victim, we need to further translate the physical addresses into more meaningful virtual addresses.
In general, a root-privileged attacker has multiple ways of obtaining the physical-to-virtual mappings: either by parsing the proc file {\tt /proc/[PID]/pagemap} (assuming Linux as the OS), or using a modified driver.
However, paging in an enclave is controlled by the SGX driver, and the vanilla driver forbids poking the physical-to-virtual mappings through the proc file system.
Nevertheless, the attack can still modify the SGX driver to retrieve the mappings, and this is what we do.  

Hence, we print the virtual-to-physical mappings in the dmesg log and ship the log together with the memory traces.
During our offline analysis, we use the dmesg log as an input to the attack script.
The dmesg log also contains system timings of paging,
and can be further calibrated to the timestamps of the collected traces.
Because paging in an enclave needs to copy the whole pages in and out of the EPC
a sequential access pattern of a whole or partial page will appear in the memory traces. After calibration, we successfully translate all the physical addresses to virtual addresses.

\section{Attack Examples}
\label{sec:example}

We show how \scheme exploits two example applications: (1) spell checking of a confidential document using {\it Hunspell}, and (2) email indexing cache using {\it Memcached}.

\subsection{Hunspell}

Hunspell is an open-source spell checker library widely used by LibreOffice, Chrome, Firefox and so on~\cite{hunspell}.
The controlled-channel attack~\cite{controlled} has shown that Hunspell is exploitable by page-granularity access patterns, which motivated us to use it as the first target of \scheme{}.
We make the same assumptions as described in \cite{controlled}; the attacker tries to infer the contents of a confidential document owned by a victim while Hunspell is spell-checking.
The attacker knows the language of the document, and therefore can also obtain the same dictionary, which is publicly available.

\begin{figure}[bt]
    \lstset{language=C,
      frame=single, showspaces=false, showtabs=false, breaklines=true,
      showstringspaces=false, breakatwhitespace=true,
      numbers=left, numberstyle=\scriptsize, stepnumber=1, frame=none, xleftmargin=2em,
      commentstyle=\color{red}, keywordstyle=\color{blue},
      columns=flexible, basicstyle=\ttfamily\footnotesize,
      escapeinside={<@}{@>}
    }
    \begin{lstlisting}
// add a word to the hash table
int HashMgr::add_word(const std::string& word) {
  struct hentry* hp = (void*) malloc(sizeof(struct hentry) + word->size());
  struct hentry* dp = tableptr[i];   // Populate hp 
  while (dp->next != NULL) {
    if (strcmp(hp->word, dp->word) == 0) {
      free(hp); return 0;
    }
    dp = dp->next;
  }
  dp->next = hp;
  return 0;
}
// lookup a word in the hash table
struct hentry* HashMgr::lookup(const char* word) {
  struct hentry* dp;
  if (tableptr) {
    dp = tableptr[hash(word)];
    for (; dp != NULL; dp = dp->next) {
      if (strcmp(word, dp->word) == 0) return dp;
    }
  }
  return NULL;
}
    \end{lstlisting}
    \caption{The Hunspell code which leaks access patterns with controlled-channel attacks and \scheme{}.}
    \label{fig:hunspell-code}
\end{figure}

The side-channel attacks on Hunspell are based on observing
the access patterns for searching words in a hash table created from the dictionary.
A simplified version of the vulnerable code is shown
in Figure~\ref{fig:hunspell-code}.
The Hunspell execution starts with reading the dictionary file and inserting the words into the hash table by calling {\tt HashMap::add\_word()}.
For each word from the dictionary,
{\tt HashMap::add\_word()} allocates a {\tt hentry} node and inserts it to the end of the linked list in the corresponding hash bucket.
Then, Hunspell reads the words for spell-checking and calls {\tt HashMap::lookup()} to search the words in the hash table.
Both {\tt HashMap::add\_word()} and {\tt HashMap::lookup()} leak the hash bucket of the word currently being inserted or searched,
and all the {\tt hentry} nodes before the word is found in the linked list.

\begin{figure}[bt]
    \centering
    \includegraphics[width=\linewidth]{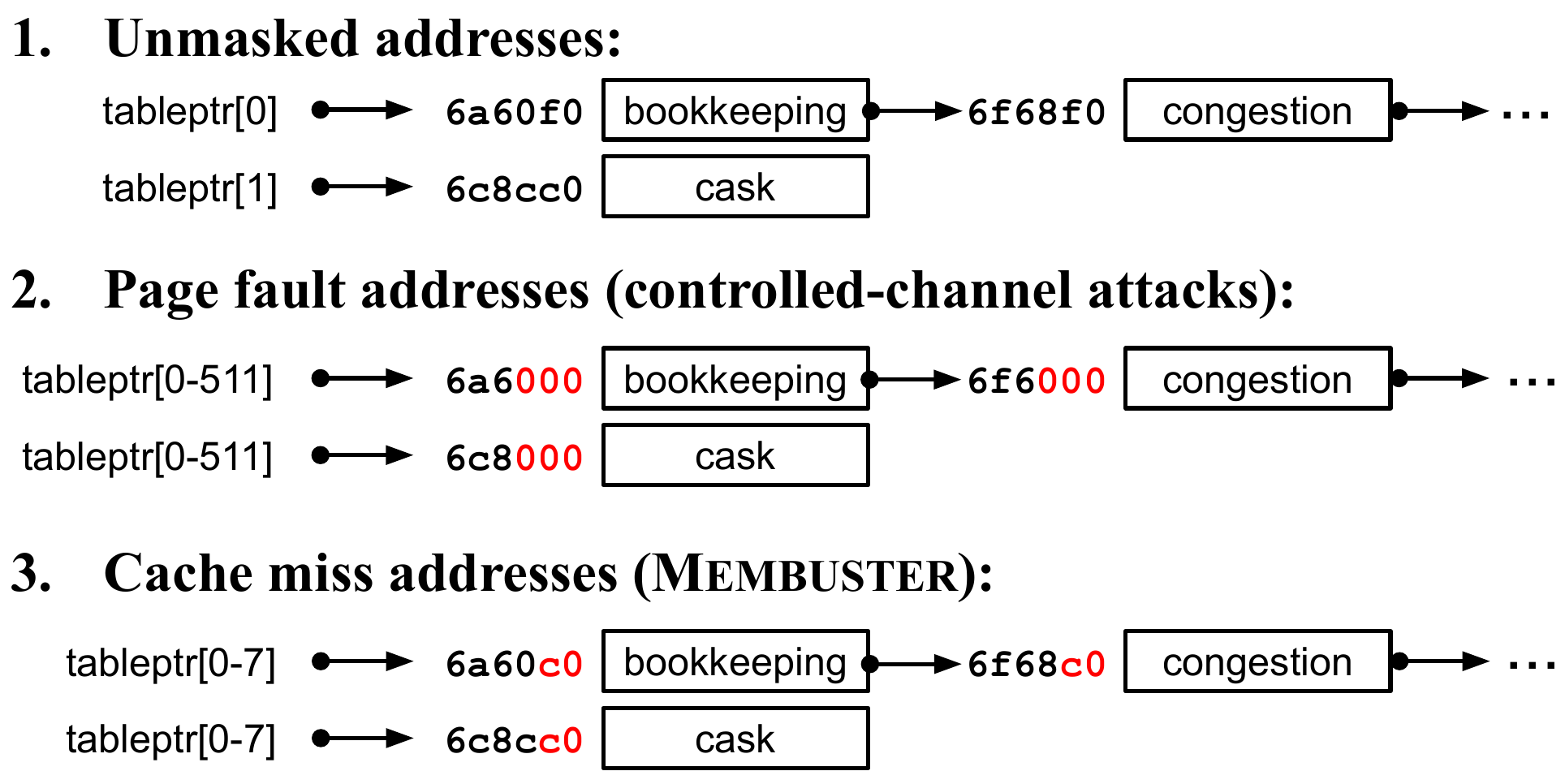}
    \caption{Observerable address patterns in Hunspell by different attacks.
    Controlled-channel attacks only see page-fault addresses without the lower 12 bits, whereas \scheme{} can see LLC-miss addresses without the lower 6 bits.}
    \label{fig:hunspell-hash-table}
\end{figure}

The controlled-channel attack leaks different access patterns from those that we observe on our memory bus attack,
as the example shown in Figure~\ref{fig:hunspell-hash-table}.
Controlled-channel attacks leak access patterns through page fault addresses, which are masked by SGX in the lower 12 bits.
However, for applications like Hunspell, controlled-channel attacks can use sequences of page fault addresses to infer more fine-grained access patterns within a page. For example, although the nodes for {\tt bookkeeping} and {\tt booklet} are on the same page, the controlled-channel attacks can differentiate the accesses by the page addresses accessed before reading the nodes.

On the other hand, our memory bus channel can leak the addresses of each cache line being read from and written back to DRAMs, making the attacks more fine-grained than controlled-channel attacks.
The attacks can differentiate the access patterns
based on the addresses of each node accessed during lookups, instead of inferring through the address sequences.
The granularity of memory bus attacks makes it possible
to extract sensitive information even if the access patterns are partially lost due to caching.

\subsection{Memcached}

Memcached~\cite{memcached} is an in-memory key-value database, which is generally used to speed up various server applications by caching the database.
Memcached is used in various services such as Facebook~\cite{memcached-facebook} and YouTube~\cite{memcached-youtube}.
In this example, we assume that Memcached runs in an SGX enclave, as part of a larger secure system (e.g., secure mail server).

\begin{figure}[t!]
    \centering
    \includegraphics[width=0.9\linewidth]{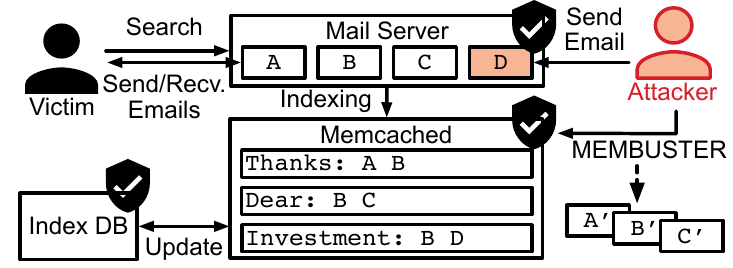}
    \caption{An example attack scenario where a mail server uses Memcached as an index database. \texttt{A}, \texttt{B}, \texttt{C} and \texttt{D} are the emails.}
    \label{fig:memcached-example}
\end{figure}

We consider the scenario discussed by Zhang~\etal~\cite{file-injection}, where a mail server indexes the keywords in each of the emails and the attacker can inject an arbitrary email to the victim's inbox by simply sending an email to the victim.
As shown in Figure~\ref{fig:memcached-example}, we assume that the index data is stored in Memcached running in an SGX enclave.
Since the attacker owns the machine, she can also perform \scheme{} by observing the memory bus.
The attacker's goal is to use his abilities to reveal the victim's secret emails \texttt{A}, \texttt{B}, and \texttt{C}.

Memcached does not have any data-dependent control flow, but the attacker can use the memory bus side channel to infer the query sent to Memcached.
Memcached stores all keys in a single hash table \texttt{primary\_hashtable} defined in \texttt{assoc.c} using the Murmur3 hash of a key as an index.
Each entry of the hash table is linearly indexed by the Murmur3 hash of the key.
Thus Memcached will access an address within the hash table whenever it searches for a key.
By observing the address, the attacker can infer the hash of the key.

Memcached dynamically allocates the hash table at the beginning of the application.
The attacker can easily find out the address of the hash table
by sending a malicious email to make Memcached access the hash table.
For example in Figure~\ref{fig:memcached-example}, the attacker sends an email \texttt{D} which contains a word "Investment".
Memcached accesses the entry, and the attacker observes the address.
Since the attacker already knows the hash value of the key, she can easily find out the address of the hash table.

Next, the attacker keeps observing the memory accesses within the hash table.
Once the attacker figures out the hash table address, she can reveal the hash values of the query, by observing the virtual addresses accessed by Memcached.
To match the hash values with words, the attacker pre-computes some natural words and creates a hash-to-word mapping.
Even though hashes can conflict, we show that the attacker can recover most of the words by just picking a most-common word based on the statistics.

\section{Increasing Critical Cache Misses}
\label{sec:improvement}

As previously discussed, the basic attack model of \scheme{}
can observe memory transactions with cache-line granularity when the memory transactions cause cache misses in the last-level cache (LLC).
Such an attack model is weakened in a modern processor with a large LLC ranging from 4~\mb{} to 64~\mb{}, causing only a small fraction of memory transactions to be observable on the DRAM bus.

In this section, we introduce techniques to increase cache misses of the target enclaves.
In a realistic scenario, an attacker only cares about increasing the cache misses within the virtual address range which leaks the side-channel information.
Take the attack on Hunspell for example, the attacker only needs to observe the access on the nodes which store the dictionary words.
We called a memory address as \textit{critical} if the address is useful for the attack.
Our goal is to increase the cache misses on critical addresses, to improve the success rate of the \scheme{} attack.

\subsection{Can We Disable Caching?}
\label{sec:improvement:whynotdisable}

A simple solution to increase cache misses is to disable caching in the processor.
On x86, entire cacheability can be disabled by enabling the CD bit and disabling the NW bit in the control register CR0 (\cite{intel-manual}, Section 11.5.3).
Some architectures allow disabling caching for a specific address range,
primarily for serving uncacheable DMA requests or memory-mapped I/Os.
For instance, on x86, users can use the Memory Type Range Register (MTRR) to change the cacheability of a physical memory range.
Newer Intel processors
also support page attribute table (PAT) to manage page cacheability with the attribute field in page table entries.

However, besides disabling the entire cacheability, neither MTRR or PAT can overwrite the cacheability of SGX's processor-reserved memory (PRM)~\cite{sgx-manual}.
The cacheability of PRM is specifically controlled 
by a special register called Processor-Reserved Memory Range Register (PRMRR),
which can be only written by BIOS during booting. 
Since there is no proprietary BIOS that allows the user to modify PRMRR,
the attacker effectively has no way to change the cacheability of the encrypted memory.
However, since the BIOS is untrusted in the threat model of SGX, in theory, 
one can 
reverse-engineer the existing BIOS or build a custom BIOS to overwrite PRMRR. 
We do not choose this route because disabling cacheability will incur significant slowdown, 
making the attack easy to detect by the victim.

\subsection{Critical Page Whitelisting}
\label{sec:improvements:pinning}

We observed that after paging (swapping), memory access
in the swapped pages becomes unobservable to the attacker.
Such a phenomenon is common for SGX since SGX has to rely on the OS to swap pages in and out of the EPC.
Both swap-in and swap-out causes the page to be loaded into the cache hierarchy (LLC, L2, and L1-D caches), because the SGX instructions for swap-in and swap-out, i.e., {\tt eldu} and {\tt ewb}, require re-encrypting the page from/to a regular physical page~\cite{sgx-manual}.
After the instructions, the cache lines stay in the cache hierarchy until being evicted by other memory access.
Currently, an Intel CPU with SGX only has up to 93.5MB in the EPC,
making paging the primary obstacle to observing critical transactions on the memory bus.

On the other hand, paging also complicates the virtual-to-physical address translation,
as the mappings can change midst execution. We observe certain patterns in the memory bus log to identify the paging events. However, these patterns can also become unobservable if the page is recently swapped and most of the cache lines are still in the LLC.

Therefore, to eliminate the side effect of paging,
we pin the EPC pages for the critical address range, by modifying the SGX driver.
We start by identifying the critical address range of each target program. Take the Hunspell program for example. The critical memory transactions come from accessing the dictionary nodes, which are allocated through {\tt malloc()}.
For simplicity, we disable Address Space Layout Randomization (ASLR) inside the enclave (controlled by the library OS~\cite{graphene}), although we confirmed that ASLR can be defeated by identifying contiguous memory access pattern in the traces.
Next, we calculate the number of EPC pages needed
for pinning the critical pages. For a Hunspell execution using an en\_US dictionary, the total {\tt malloc()} range is 5,604~\kb{}.
Finally, we need to give the critical address range as an input to the modified SGX driver.
When the driver allocates an EPC page, it checks if the virtual address is in the critical address range and use an in-kernel flag
to indicate if the page has to be pinned.
The driver will never swap out a pinned page.

\subsection{Priming the Cache}

We explore ways to actively contaminate the caches by accessing contentious addresses.
This technique is called {\em cache priming}, which is used in the \primeprobe attack~\cite{liu15cache-attack}.
Previous work has established priming techniques for either same-core or cross-core scenarios.
Some priming techniques are restricted by CPU models,
especially since many recent CPU models have employed designs or features that raise the bar for cache-based side-channel attacks.
However, recent studies also show that, even with these defenses, attackers continue to find attack surfaces within the CPU micro-architectures, such as priming the cache directory in a non-inclusive cache~\cite{yan19attack}.

We focus on cross-core priming since same-core priming requires interrupting the enclaves using AEX or page faults.
The usage of cache priming in \scheme is distinctly different from existing cache-based side-channel attacks
since \scheme does not require resetting the state of the cache or synchronizing with the victim.
The goal of cache priming in \scheme is to simply evict the critical addresses from the cache to increase the cache misses.
Also, with \squeezing, we only have to prime the cache sets dedicated to the critical addresses.
These differences make it easy to apply multiple priming
attacks simultaneously, as long as they all eventually contribute to increasing cache misses.

\paragraph{Cross-Core Cache Priming}
\label{sec:improvement:cross-core}

We run multiple priming processes on other cores
to evict the critical cache lines from the LLC.
These processes will repeatedly access the cache sets that are shared with the critical addresses of the victim.
The attacker will start by identifying the critical addresses
and the cache sets to prime.
Then, the attacker starts the priming processes before the victim enclave, to actively evict the cache lines during execution.
Take the Hunspell attack for example.
Since its critical addresses are spread over all cache sets, the attacker needs to repeatedly prime all cache sets.
No synchronization is required between the attack processes and the victim.
We do not prime the L1 and L2 caches across cores,
but cross-core priming on private caches is demonstrated
on Intel CPUs~\cite{yan19attack}.

A potential hurdle for \priming is to obtain sufficient memory bandwidth to evict the critical cache lines.
Based on our experiments, a priming process that sequentially accesses
the LLC has around 100--200MB/s memory bandwidth.
Priming a 9MB LLC with 2,048 sets
requires about 100 milliseconds, 
which is too slow to evict the critical cache lines before the lines are accessed by the victim again.
For instance,
\hunspell accesses a word every 2 thousand DRAM cycles ($<1$ microseconds), and \memcached
accesses a word every 5 million DRAM cycles ($<2.5$ milliseconds).
We will discuss, however, how an attacker can evict all the critical cache lines within a few milliseconds
by pinpointing the priming process to target only 64--128 sets 
(See \S\ref{sec:prime+squeezing}).

\paragraph{Page-Fault Cache Priming}
Potentially, an attacker can prime the LLC, L2, and L2-D caches on the same core with the victim, by interrupting the victim periodically.
To do so, the attacker can take a similar approach to the Controlled-Channel Attack:
The attacker identifies two code pages containing code around the critical memory accesses,
and then alternatively protects the pages
to trigger page faults.
To increase cache misses, the attacker needs not to prime the cache at every page fault, but rather can prime at a low frequency.
However, such a page-fault priming technique still
causes a lot of interference and overhead to the victim, making it easy to detect~\cite{cachezoom} or to mitigate~\cite{tsgx,chen17side-channel}.
For example, priming the cache on every 10-20 page faults
incurs about $3\times$ overhead to the victim.
In addition, known countermeasures, such as T-SGX~\cite{tsgx}, can effectively prevent page faults using transactional instructions.
Therefore, we do not use this technique.

\subsection{Shrinking the Effective Cache Size}
\label{sec:improvements:squeezing}

As previously discussed, cache priming alone cannot 
create sufficient memory access bandwidth for evicting the critical cache lines in time.
Therefore, we introduce a novel technique called {\it cache squeezing}, which 
shrinks the effective cache size to incur more 
cache misses for a specific address range.
We show that the technique can be combined with non-intrusive techniques like cross-core cache priming to
make \scheme a more powerful side channel.

\begin{figure}[bt]
    \centering
    \includegraphics[width=\linewidth]{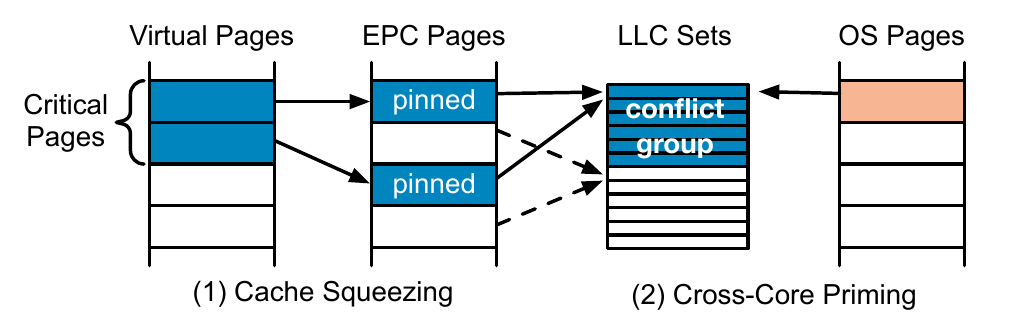}
    \caption{Techniques used to increase the cache miss rate with minimal performance overhead.}
    \label{fig:cache-squeezing-diagram}
\end{figure}

\subsubsection{Cache Squeezing}\label{s:squeezing}

As the name suggests, cache squeezing can shrink the effective cache size for a given set of critical pages.
By squeezing the cache that an enclave can use, the attacker can incur both {\em conflict misses} and {\em capacity misses} on LLC, therefore becoming able to observe more cache misses on the bus.

In modern processors, the L2 cache and LLC are physically-indexed.
The lowest 6 bits of the physical address are omitted, given that each cache line is 64 bytes.
The next $s$ lower bits are taken as the {\em set index}. 
Each set then consists of $W$ ways
to store multiple cache lines of the same set index.
For an enclave,
an OS-level attacker can control the physical pages that are mapped to the enclave's virtual pages.
This allows the attacker to manipulate the physical frame number (PFN) of each virtual address of the enclave,
and subsequently,
the higher $s-(12-6)=s-6$ bits of the set index.

Figure~\ref{fig:cache-squeezing-diagram}(1) shows how cache squeezing works in combination with page pinning.
The attacker first defines the critical addresses of the victim, then maps these pages to EPC pages that share the minimum amount of cache sets.
This technique requires cache pinning so that
these pages will never be swapped out from the EPC.
Since the OS only controls the higher $s-6$ bits of the set indices, the smallest group of physical pages that will evict each other share exactly $2^6=64$ sets.
We called such a group of physical pages
a {\em conflict group}.
Since the maximum size of EPC is 93.5~\mb{}, the entire cache can be partitioned to $2^{s-6}$ conflict groups where each conflict group can accommodate 93.5~\mb{}/4~\kb{}/$2^{s-6}$ EPC pages.
In our experiment, $s=11$ (2048 sets) and $W=12$, so each conflict group can accommodate at most 748 pages (2,992~\kb{}).
The critical address range of Hunspell, for example, is the whole {\tt malloc()} space, which is 5,604~\kb{} and thus requires two conflict groups.
Finally, the attacker gives the critical address range to a modified SGX driver,
which will only map physical pages from the selected conflict groups to any critical virtual address.

Using cache squeezing to increase cache misses has many benefits. First of all, it does not require interrupting the victim enclaves,
nor does it need to incur more memory accesses in the background. All memory accesses used to push cache lines out of the L2 cache and LLC are legitimate accesses from the victim enclaves.
Therefore, cache partitioning cannot defeat cache squeezing because there is no cross-context cache sharing.
In fact, way-partitioning features such as Intel CAT~\cite{intel-cat} can be exploited to further shrink the effective cache sizes in combination with cache squeezing.

\subsubsection{Cross-Core Priming with Cache Squeezing}
\label{sec:prime+squeezing}
As we mentioned in \S~\ref{sec:improvement:cross-core},
cross-core cache priming
may not have sufficient bandwidth to evict the critical cache lines in time.
However, we found that cache squeezing
makes the priming more effective by shrinking the effective cache size.
Instead of priming all the cache sets, the attacker now only has to prime the sets of the targeted conflict groups containing the critical addresses (Figure~\ref{fig:cache-squeezing-diagram}(2)).
Each group of 64 cache sets contains $W\times 4$KB, allowing the priming process to evict the part of cache
within a millisecond. The priming process can run in parallel and does not affect the victim execution
except causing cache contention.

\subsubsection{Limitation}
\label{sec:improvements:squeezing:limitation}
Although cache squeezing can increase the cache misses among critical addresses, 
it could be less effective if the victim has only a few critical addresses or a small memory footprint.
If the critical addresses can only fill a small part of a conflict group ($W\times4$ KB),
the victim enclave may not be able to cause enough cache misses to benefit the attacker.
For example, \memcached only has $2$ MB ($500$ pages) of the critical address range.
To fill all of the 748 pages, we identify
the top 248 frequently-accessed pages (in addition to the critical addresses) through simulation,
and assign these extra pages to the same conflict group.

Note that the LLC of a modern CPU usually has a {\em cache slice} feature that distributes the addresses across multiple cache banks 
using an undocumented, model-specific mapping function. 
Reverse-engineering the slicing function
of the target CPU is useful
for further reducing the effective cache space for an enclave if the enclave has a smaller memory footprint.
Reverse-engineering of slicing functions
is already explored by prior papers~\cite{yan19attack}, so we will not discuss this technique in this paper.

One can detect the cache squeezing by testing
if critical addresses are mapped in an adversarial way.
Since the enclave is not aware of physical address mappings by itself, it needs to experimentally
detect such mapping by accessing the addresses and measure latency.
However, we claim that it is challenging because 
(1) the victim needs to know the critical address range to detect the mapping,
and (2) the enclave cannot tell if the mapping was accidental or intentional.

\subsubsection{Implementation}

\begin{figure}[bt]
    \centering
    \includegraphics[width=\linewidth]{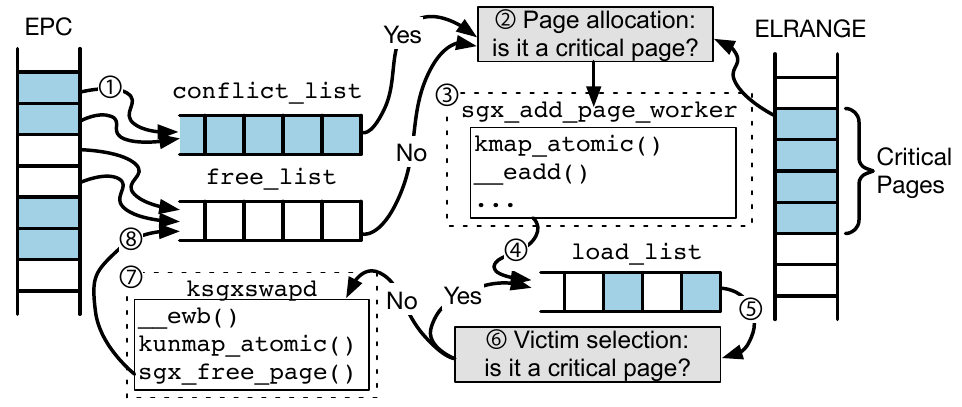}
    \caption{Implementation of \pinning{} and cache squeezing in a modified SGX driver. To ensure no swapping in the sensitive memory range, EPC pages are set aside in a separate queue. The attackers can further select the EPC pages based on set indexes or other logistics.}
    \label{fig:cache-squeezing}
\end{figure}

We use a modified SGX driver to implement both \pinning{} and cache squeezing as shown in Figure~\ref{fig:cache-squeezing}.
The driver accepts parameters to specify a sensitive range within the victim application, and calculates how many conflict groups are required for the attack.
\textcircled{\small{1}} When the driver initializes, it inserts conflicting EPC pages to a separate queue (i.e., {\tt conflict\_list}). 
\textcircled{\small{2}} When adding enclave pages, the driver checks if the virtual page number is in the critical address range.
\textcircled{\small{3}} The driver maps the critical pages to pages popped from {\tt conflict\_list}.
\textcircled{\small{4}} All of the mapped pages are added to the list of loaded pages ({\tt load\_list}).
\textcircled{\small{5}} When the driver needs to evict an EPC page, it searches the victim from the list of loaded pages.
\textcircled{\small{6}} If the selected page is a critical page, it searches again.
\textcircled{\small{7}} Only non-critical pages are evicted and the enclave continues to run. 
Other enclaves are not affected by the modification and can function normal with marginal overheads.

Our change to the SGX driver contains only 290 lines. The SGX driver uses the {\tt fault} operation in {vm\_operations\_struct} to handle EPC paging.
We use a customized {\tt fault} function, which checks the faulting virtual addresses of the enclave and then applies different paging strategies to critical and non-critical addresses. We hard-code the range of critical addresses for each application and thus require switching the drivers for a different target. Potentially, the driver can export an API to the attackers for specifying the critical addresses. Our driver also only supports one single victim enclave at a time. However, we can extend the driver to target multiple enclaves simultaneously as long as the total memory usage can fit into the EPC (required for pinning).

\section{Extracting Sensitive Access Patterns}
\label{sec:matching}

OS techniques including \pinning, \squeezing, and \priming effectively increase the cache misses on the cache misses on critical addresses.
However, the traces collected from the memory bus are still full of noise and contain no marker for splitting the critical memory accesses
into iterations.
Unlike controlled-channel attacks, \scheme{} cannot rely on repeated code addresses (e.g., from a loop) to mark and then split the critical accesses
because these code addresses tend to be accessed too frequently to be evicted by our techniques.
Therefore, the attacker needs to deeply analyze the memory traces offline to distill the sensitive information.

To extract the sensitive access patterns, we identify four techniques for filtering the critical memory addresses and matching with a known oracle for the target application:
(1) offline simulation; (2) searching the beginning of sensitive accesses; (3) fuzzy pattern matching, and (4) exploiting cache prefetching.
We use the two examples
to explain how to analyze memory bus traces.

\subsection{Offline Simulation}

Side-channel attacks often require attackers to have some knowledge about the behaviors of the victim.
For example, the controlled-channel attack on Hunspell requires the attacker to extract the virtual page addresses of the linked list nodes of each dictionary word, during an online {\em training phase} while attacking the victim.
However, \scheme{} cannot perform online training with the victim
as the analysis of the memory traces is performed offline.
Instead, the attacker needs to generate an oracle of the victim behavior, using offline simulation of the target application.

We observe that, for each application, we can use a deterministic oracle,
given that users have adopted some 
publicly available data (e.g., the {\tt en\_US} dictionary).
For example, during the simulation,
we run a modified Hunspell in an enclave, which prints out the indexes and the addresses of linked list nodes visited for each word.
Then, we reuse the output as an oracle,
to be used in analyzing any traces based on the same {\t en\_US} dictionary.
We assume that there are only a finite amount of English dictionaries
in the world.

As discussed earlier, ASLR in the enclaves does not invalidate an oracle,
since ASLR can be easily defeated by observing the specific patterns related to binary loading.
The addresses in the oracle can simply be shifted by a certain offset to be usable again.

\subsection{Searching Sensitive Accesses}

Finding the first sensitive access is critical for deciding where to start matching access patterns.
Note that not all accesses to the critical addresses are sensitive.
For Hunspell, allocating nodes for each word emits a long sequence of monotonically increasing virtual addresses that can be used to identify the sensitive addresses. 
We match the virtual addresses to the oracle, to find the {\em longest increasing subsequence (LIS)} of addresses as accessed in the dictionary order.
After finding the LIS, the next critical access is the beginning of the sensitive addresses.

\subsection{Fuzzy Pattern Matching}

In \scheme{}, we observe that a part of memory addresses in a sensitive access pattern is likely to be missing due to caching.
Even with \squeezing and \priming, it is almost impossible to force page misses on every critical memory access.
Therefore, to analyze lossy traces,
we use fuzzy pattern matching to flexibly match the traces with only parts of access patterns.
As long as at least one or a few accesses of a pattern cause LLC misses,
we can identify the pattern as a possible result for recovery.

In fuzzy pattern matching, one address may be parsed as different access patterns of the victim for two reasons.
First, within a data structure such as a linked list or a tree,
the same address (an inner node) may be accessed while traversing or searching other nodes.
Second, a cache line may contain multiple nodes and thus can be accessed when visiting one of the nodes.
For either of the reasons, a single memory trace may be accounted for multiple possible access patterns in the oracle.

We use a simple strategy to select the best interpretation for a set of memory traces.
We assign a score to each possibility based on
how {\em complete} the traces
have matched with an access pattern in the oracle.
For the addresses of a tree or a linked list, we assign lower scores to the root and the first few nodes and assign higher scores to nodes that are closer to leaves or the end of the list.
By collecting the top-ranking interpretations of the memory traces, an attacker can generate a list of the most probable options of the target secret.
Potentially, a grammar checker or any semantic-based heuristic can help to validate or to rank the recovery results.
We leave the exercise of applying more context-aware heuristics for  future work.

\subsection{Exploiting Cache Prefetching}

Finally, we observe that the cache prefetching features of CPUs can help increase the accuracy of the attack.
For example, a recent Intel CPU includes {\em Next-line Prefetcher} and {\em 128-byte Spatial Prefetcher}.
The Next-line Prefetcher, belonging to the L2 cache, will preload the cache line next to the one that is currently accessed.
The 128-bit Spatial Prefetcher, which also belongs to the L2 cache, prefetches the pairing cache line that completes the accessed cache line to a 128-byte aligned chunk into the LLC.
Both prefetchers increase the number of memory accesses relevant to the secret data.
Therefore, we expand the range of pattern matching based on our knowledge of cache prefetching,
including extending the addresses representing each secret by 64 bytes,
both backward and forward.
As a result, even if the CPU has cached a line,
the prefetched lines may still cause cache misses and be observed on the memory bus.

Other cache prefetchers such as {\em Stream Prefetcher} can monitor an ascending or descending sequence of addresses from the L1 or L2 cache and can prefetch up to 20 cache lines ahead of the loaded address.
Such a prefetcher generally will not improve the accuracy of the pattern matching. However, these prefetchers can cause space pressure to caches, making \squeezing more effective.

\section{Evaluation}
\label{sec:eval}

In this section we present the evaluation results of the \scheme{} attack, based on the two vulnerable applications described in \S\ref{sec:example}. 
The evaluation mainly answers the following questions
regarding the \scheme{} attacks:
\begin{compactitem}
	
\item How accurate can \scheme{} extract the secrets from applications that are vulnerable to such an attack?

\item How do the attack techniques of \scheme{} impact the attack accuracy?

\item How much slowdown (or interference) the various techniques will incur on the applications?

\item What is the limitation of \scheme{}?

\item How sensitive are the attack results of \scheme{} to the last-level cache (LLC) size of the target CPU?
	
\end{compactitem}

We evaluate the \scheme{} attack in various settings: (1) the basic attack without any techniques ({\bf None}); (2) the optimized attack with \squeezing ({\bf SQ}); (3) the optimized attack with \squeezing combined with cross-core cache priming ({\bf SQ+PR}).

\subsection{Experiment Setup}

In this section, we describe the experimental setup of the \scheme{} attack. We use both physical and simulated experiments to evaluate the effectiveness of \scheme{}.

\begin{table}[t!]
    \footnotesize
    \renewcommand{\arraystretch}{1.1}
    \setlength{\tabcolsep}{0.5em}
    \centering
    \begin{tabular}{ll}
    \hline
        \multicolumn{2}{c}{\cellcolor{gray!55}\bf CPU}  \\
    \hline
    Model    & Intel i5-8400 (Coffee lake) \\
    LLC Size & 9~\mb{} \\
    LLC \# Slice & 6 Slices \\
    LLC \# Associativity & 12-way set associative \\
    LLC \# Sets & 2048 \\
    \hline
        \multicolumn{2}{c}{\cellcolor{gray!55}\bf Memory}  \\
    \hline
    DIMM Type & DDR4-2400 UDIMM (Non-ECC)\\
    Capacity &  8~\gb{} \\
    Channel/Rank/Bank/Row    & 1/1/16/65536 \\
    Page Size   & 8~\kb{} (1~\kb{}/package)\\
    Max Bus Frequency & 1200~MHz \\
    \hline
    \end{tabular}
    \caption{Hardware specification for the experiment}
    \label{tab:experiment-setup}
\end{table}

\subsubsection{Physical Experiment}
\label{sec:physical_experiment}
\noindent {\bf Hardware Setup. } The hardware setup we used for the experiment is shown in Table~\ref{tab:experiment-setup}. 
We use a machine equipped with an Intel SGX CPU.
In the machine, we connect
the DIMM to a signal analyzer via a DIMM interposer.
We configure BIOS to slightly increase the DRAM supply voltage to offset the voltage drop caused by the interposer.
The bus frequency is set to 1066 MHz, so the bandwidth of the analyzer is 3.97 \gib/s. With a 64 \gib{} acquisition depth, we can log the memory bus for up to $\sim 16$ seconds. 
All of our experiments have finished in a few seconds,
and thus the acquisition depth is sufficient for logging all the memory requests. 
To achieve a wider time window, the attacker can choose an analyzer which can filter the requests by addresses~\cite{protocol-analyzer2},
or which has a higher acquisition depth~\cite{jki-analyzer}.

\noindent {\bf Victim Setup. } The victim machine is running Ubuntu 16.04 and Linux kernel 4.4. 
To execute the victim applications inside enclaves, 
we use Graphene-SGX~\cite{graphene} to run unmodified binaries with SGX.
The victim may also choose other frameworks~\cite{scone} or port the applications with the SDK~\cite{intel-sgx-sdk},
but the choices of the frameworks do not eliminate the patterns
since they do not change the program logic of the victim.

\noindent {\bf Sample Size.} We collaborate with SK Hynix to use its proprietary analyzer for the experiments. Due to the limited access to the device, we run the attack only \textit{once} for each setting. 
However, we were able to successfully perform the attack despite the small sample size because the results match well with our expectations learned from the simulation.

\subsubsection{Microarchitectural Simulation}
\label{section:eval-simulation}

We also implemented a software simulator to simulate the attack prior to an actual attack because the hardware setup requires costly devices.
We use the simulator for exploring the attack and getting preliminary results.
The results are then cross-validated with the results from the actual hardware setup, to verify the functional correctness of the simulation.
The attacker can also use the same strategy to save the expenses for renting the devices.
We modify QEMU~\cite{qemu}, a machine emulator, to trace all the physical memory accesses of the guest.
To capture cache misses, we make QEMU emits all the memory requests
to a cache simulator we integrated from Spike~\cite{spike}.
The cache simulation does not 
implement any cycle-accurate hardware model as well as cache slicing and pseudo-LRU replacement.
However, the simulation was sufficiently faithful for developing the attack scripts to analyze the real memory traces.

\subsubsection{Enclave Simulation}

We also simulate an enclave environment without memory encryption, using a modified Graphene-SGX library OS and a dummy SGX driver.
We consider simulating Intel's Memory Encryption Engine (MEE) unnecessary because
MEE does not affect the memory addresses accessed within the EPC.
MEE generates additional access patterns for the integrity tree or EPC metadata, both of which are stored in the Processor Reserved Memory outside the EPC. Our attack does not rely on any access pattern outside the EPC.  

The modified Graphene-SGX library OS and the dummy SGX driver
primarily simulate the transition in and out of the enclave and the paging of enclave memory, to generate similar memory access patterns as observed on the memory bus.
For simulating enclave entry and exit, we modify the user-tier SGX instructions, {\tt EENTER} and {\tt EEXIT}, in the Graphene-SGX runtime, to directly jump to addresses that are originally given as the enclave entry.
We also simulate the AEX.

For simulating EPC paging, we modified the SGX driver to replace the system-tier SGX instructions, including the {\tt ELDU} and {\tt EWB} instructions,
which swap and re-encrypt pages in and out of the EPC.
We simply replace these two instructions with memory copy
without encryption.
We compare the memory traces from the real enclaves and from the simulation to confirm that the results are identical.

\subsubsection{Applications: Hunspell}

We run Hunspell v1.6.2 to evaluate the effectiveness of the \scheme{} attack.
We use a standard \texttt{en\_US} dictionary~\cite{wordlist} with two document samples: a random non-repetitive document with 10,000 words ({\bf Random}), 
and a natural-language document ``Wizard of Oz'' with 39,342 words ({\bf Wizard}).
For simplicity, we normalize the samples
based on \texttt{en\_US} dictionary, by converting non-existing words in the samples to the closet words in the dictionary.
\scheme{} does not recover words that are reported as misspelt by Hunspell. 
In addition, we disabled affix detection in Hunspell.

We use the pattern matching algorithm described in \S\ref{sec:matching} 
to recover the target document from
the DRAM traces
collected from the Hunspell program running inside the enclave.
We also enable the hardware prefetching by configuring the BIOS.
To verify the result,
we select an interpretation of the DRAM traces that is closet to the target document,
from a set of highest-ranking results
generated from our algorithm.

\subsubsection{Application: Memcached}

We run Memcached v1.5.12
as another target of the \scheme{} attack.
In this attack, the ``secrets'' are the data being looked up in the Memcached cache.
We used the Enron email dataset~\cite{enron-dataset}
as a realistic workload for Memcached.
First, we compute the 4-byte hash of each word that appears
in emails in the ``sent mail'' directory of each user.
In total, there are
about 7000 unique word entries in the dataset, which include articles and propositions.
During the {\em training phase}, assuming the attacker is monitoring a Memcached server,
the attacker can determine 
both the hash table address and the hash value of each word using the traces of a few queries.
Then, during the {\em attack phase},
the attacker monitors the memory bus traffic of an enclave-protected Memcached server receiving caching requests from an trusted email server.
The email server parses emails
from a test data set that contains randomly selected emails with around 1000 words in total.
As the Memcached server processes the caching requests 
from the email server,
the attacker can extract the words in the emails
using the \scheme{} attack.

\subsection{Effectiveness of the Attack}

\begin{figure}[t!]
	\centering
	\begin{subfigure}[t]{.55\linewidth}
	\vskip 0pt 
		\includegraphics[width=\linewidth]{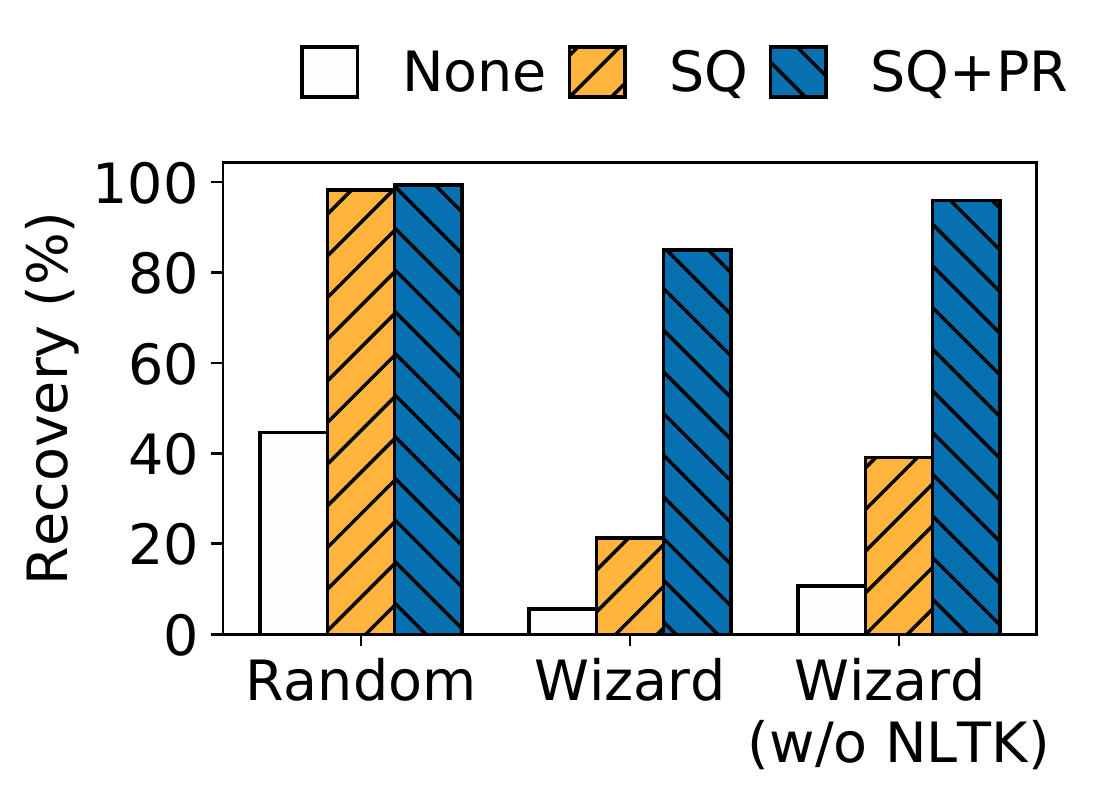}
	\end{subfigure}
	\hspace{0pt}
	\begin{subfigure}[t]{.42\linewidth}
	\vskip 0pt
		\includegraphics[width=\linewidth]{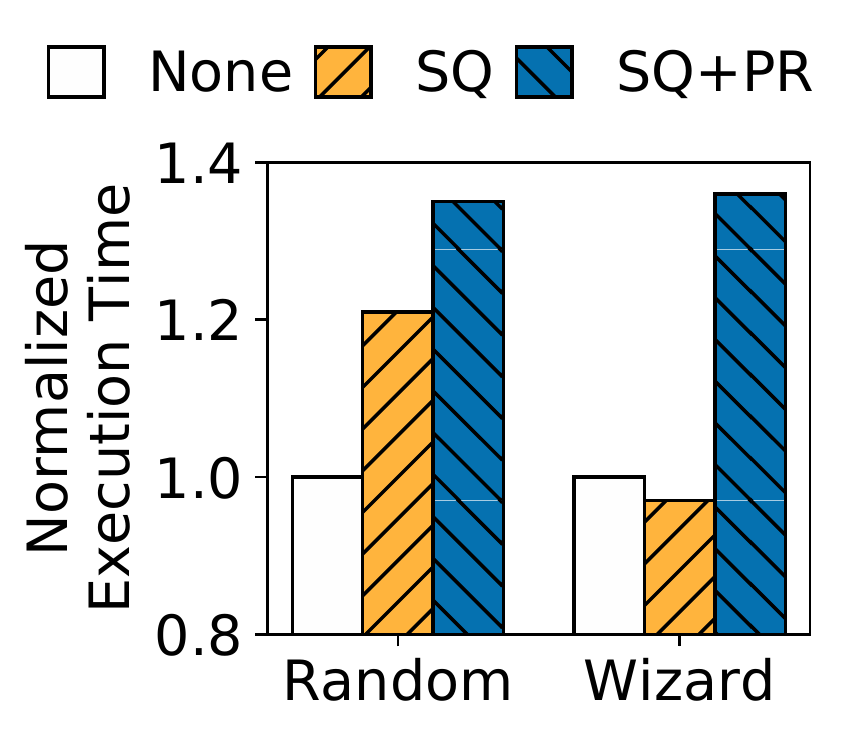}
		
	\end{subfigure}
	\caption{Hunspell document recovery rate (left) and normalized execution time (right) on two documents: Random document (Random) and Wizard of Oz (Wizard). The comparison is between without any techniques (None); with \squeezing(SQ); and with \squeezing and \priming (SQ+PR). For Wizard of Oz, we also show the recovery rate of uncommon words only (w/o NLTK).}
	\label{fig:hunspell_results}
\end{figure}

\begin{table}
	\footnotesize
	\centering
	\renewcommand{\arraystretch}{1.2}
\begin{tabular}{r|r|r}
	Technique       & Attack Accuracy         & Normalized Exec. Time  \\
	\hline
	None    &   34.1\%  &  $1.00 \times$ \\
	SQ       &    82.1\%      & $0.92 \times$\\ %
\end{tabular}
\caption{\scheme results for attacking Memcached on an SGX machine}  
\label{tab:eval_memcached}
\end{table}

\subsubsection{Data Recovery Accuracy}

Figure~\ref{fig:hunspell_results} (left) and Table~\ref{tab:eval_memcached} show the accuracy of \scheme{} for recovering the victim's data.
We measure the accuracy based on the number of words
recovered from the collected traces, compared to the number of words in the original samples.
The recovery rate is higher in a non-repetitive (Random) or high-interval access pattern (Memcached)
than in a repetitive access pattern (Wizard).
Even without any techniques (None), Memcached and Random show $34\%$ and $44\%$ recovery rates, respectively.
With \squeezing, we recover $96\%$ of the random document and $82\%$ of the Memcached query.

However, for Wizard of Oz,
None or SQ can only achieve up to $21\%$ recovery rate.
The main reason is that the document contains many repetitive words, including common words such as ``you'' and ``the'' and uncommon words such as ``Oz'' and ``scarecrow''.
The memory accesses
for these words
are likely to be cached in the LLC cache without emitting any DRAM requests.
On average, each unique word in Wizard of Oz repeats 15.5 times.
We found that without \squeezing and \priming, the attack recovers about 
0.3 occurrences of each word on average.
Even with \squeezing, the attack only recovers about 2.6 occurrences.

Since \squeezing shrinks the effective cache size for the critical addresses,
\priming becomes more efficient
by only priming the sets of the critical addresses.
We show that combining \squeezing and \priming (SQ+PR) achieves $85\%$ recovery accuracy on Wizard of Oz.

Furthermore, the attacker is most likely to need only
the {\em uncommon} words to be recovered.
To exclude common words,
we use {\em stopword}s from the NLTK dataset~\cite{nltk} which includes 179 common words (e.g., "the").
Excluding these words, \scheme{} can recover Wizard of Oz up to 95\% (Figure~\ref{fig:hunspell_results} Wizard w/o NLTK).

\subsubsection{Overhead and Interference}

We show that \scheme{} does not incur an orders-of-magnitude overhead that can be distinguishable by the victim.
Figure~\ref{fig:hunspell_results} (right)
shows the normalized execution time with different attack techniques with respect to the baseline.
In general, both \squeezing{} and \priming have a low performance impact on the victim program,
since these techniques do not interrupt the victim program.
For Hunspell, \squeezing{} causes up
to 21\% overhead to the victim, and up to 36\% if combined
with \priming.
The overheads are mainly caused by the increase of cache misses
inside the victim program.

Table~\ref{tab:eval_memcached}
also shows 
the end-to-end execution time of Memcached for processing the whole test set.
Similar to Hunspell, the basic attack incurs no overhead on Memcached.
Interestingly, \squeezing reduces the execution time by 8\% for Memcached.
We observe that, on a physical machine, \pinning{} consistently reduces the average LLC miss rate (2.9\% vs. 3.6\%) as well as the page fault rate.
Because the physical pages of Memcached's hash table are pinned inside the enclave, and thus never get swapped out from the EPC.
Thereby, within the hash table, there is no expensive paging and context switching cost that generally plagues enclave execution.

\subsubsection{Scalability on \# of Ways}

We simulated the attack on our simulation environment to show the scalability of \scheme.
We fixed the number of sets $s=2048$ that most Intel CPUs choose to have.
Since we did not simulate the LLC slices, we increased the size of the cache by increasing the number of ways, $W$.
To clarify, increasing the number of ways does not reflect the actual behavior of LLC with multiple slices.
Even if the LLC has multiple slices, each cache line will compete with $W$ other cache lines. 
Thus, increasing $W$ makes the attack much harder, by reducing the chance of eviction of critical addresses.
Note that a typical $W$ value is between 4 and 16.

\begin{figure}[t!]
	\centering
	\begin{subfigure}[t]{0.9\linewidth}
		\includegraphics[width=\linewidth]{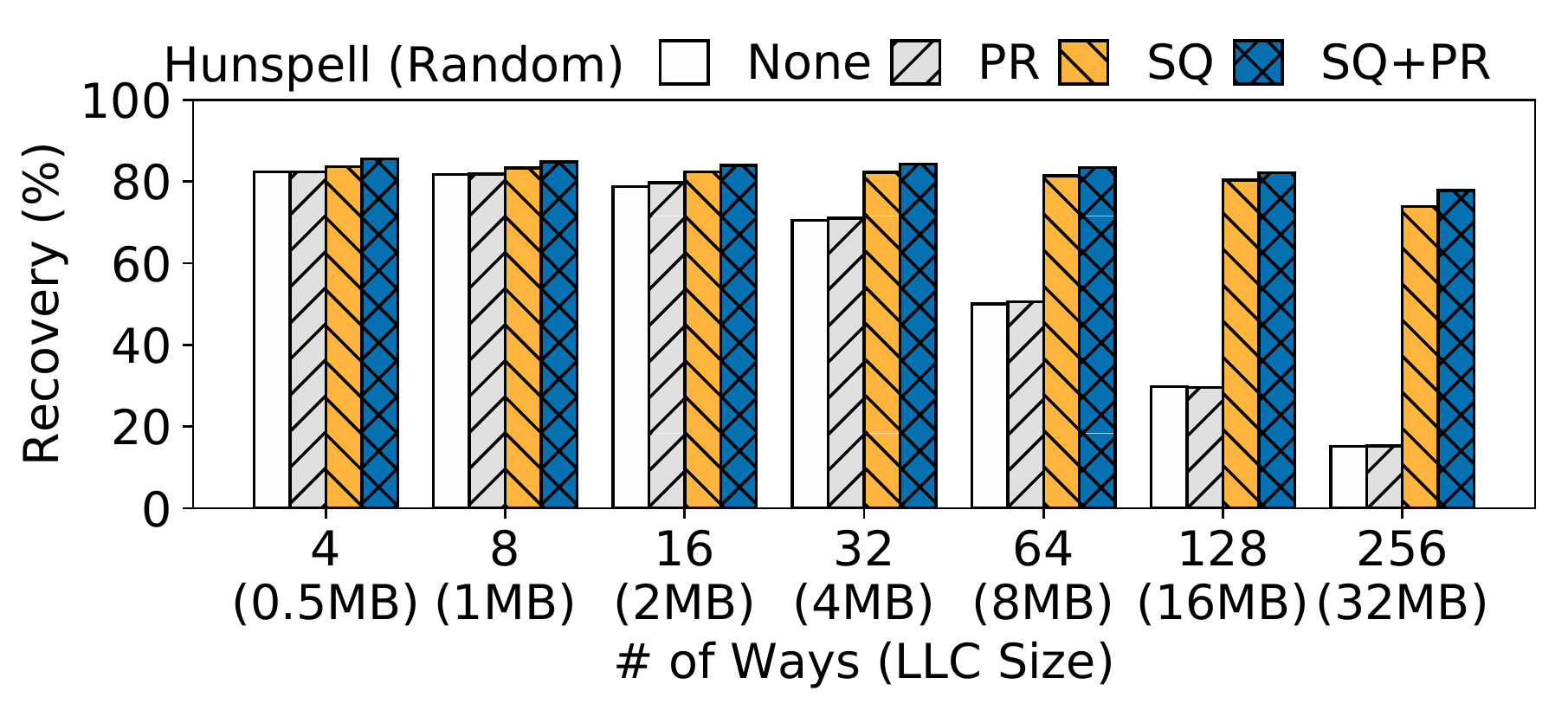}
	\end{subfigure}
	\begin{subfigure}[t]{0.9\linewidth}
		\includegraphics[width=\linewidth]{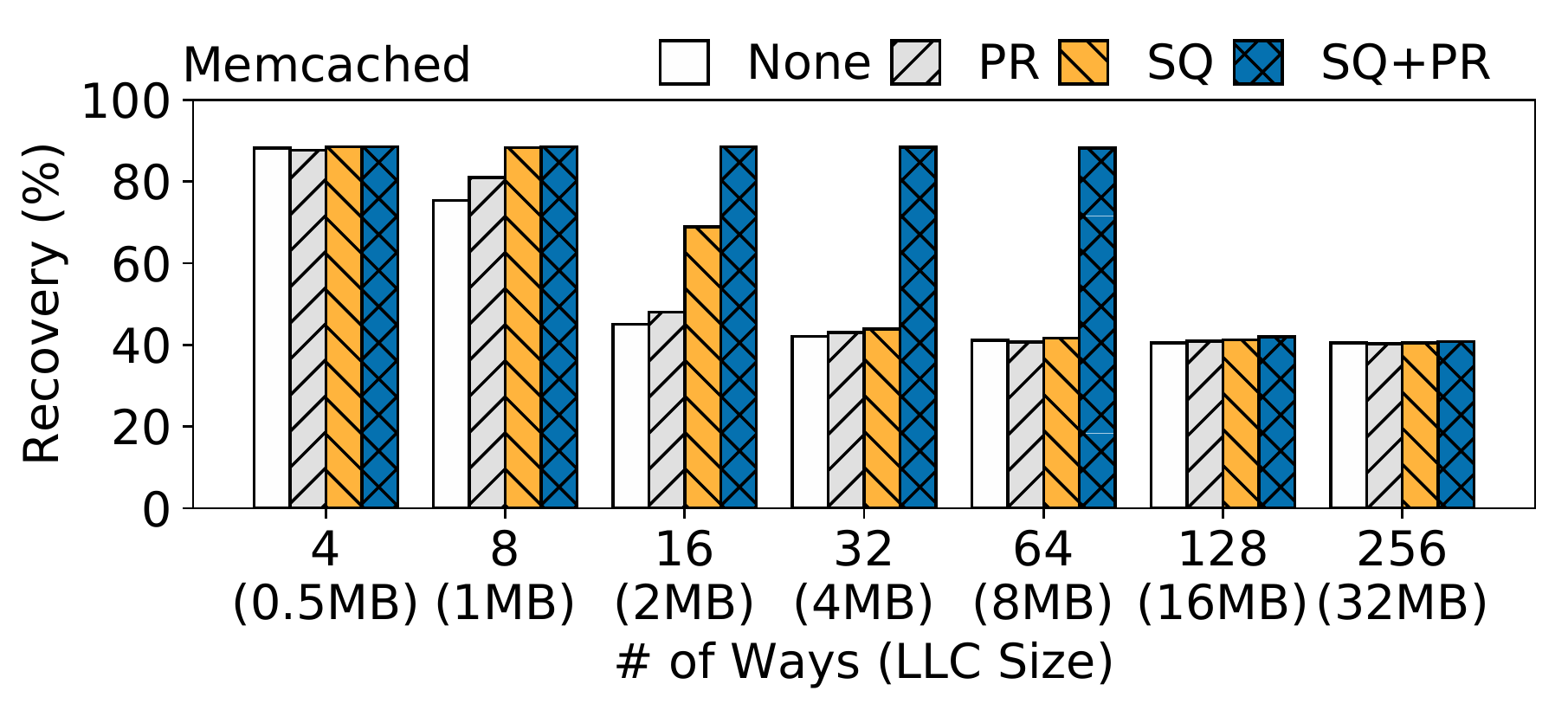}
	\end{subfigure}
	\caption{Simulation results of the attack on Hunspell (top) and Memcached (bottom).}
	\label{fig:eval_sim}
\end{figure}

As shown in Figure~\ref{fig:eval_sim}, \squeezing makes \priming much more effective in general by reducing 
the effective cache size.
\Squeezing was more scalable on Hunspell than Memcached, because Hunspell has a larger critical address range.
With $W=64$, \scheme{} recovered up to 83\% of the random document in Hunspell and 88\% of the emails in Memcached when both cache squeezing and \priming have been used.
Even assuming an unrealistic number of ways $W=256$, which results in 32~\mb{} of LLC, the attack accuracy was 77\% and 40\% respectively.

\subsection{Per-Application Detailed Analysis}

\subsubsection{Hunspell: Advantage of Cache Prefetching}
\label{sec:eval:per_app_hunspell}

\begin{figure}[t!]
	\centering
	\includegraphics[width=\linewidth]{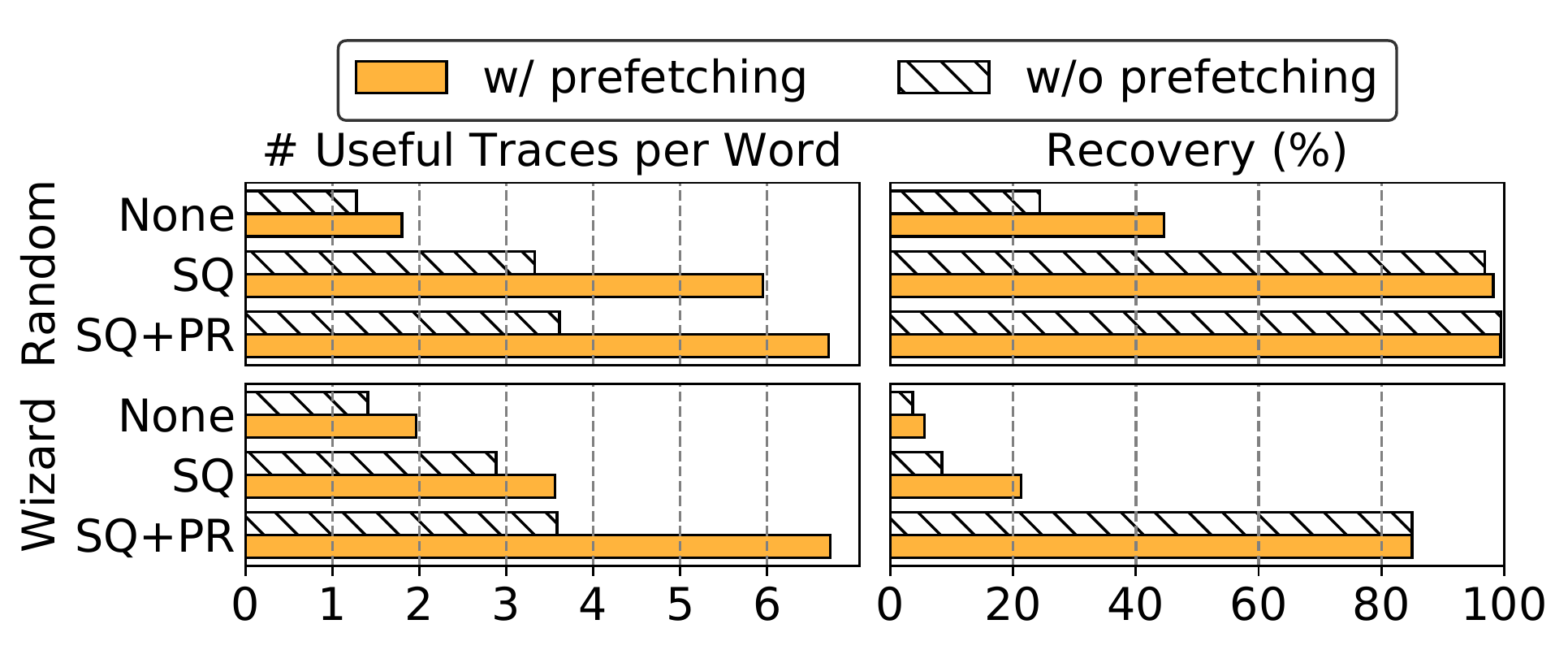}
	\caption{The number of useful traces per word and the document recovery rate for each experiment.
	We compare the cases with or without the hardware prefetcher.}
	\label{fig:hunspell_confidence}
\end{figure}

We also show the advantage of exploiting cache prefetching for \scheme.
For Hunspell, the attacker recovers each word based on multiple memory accesses.
If the attacker observes more traces relevant to each word,
recovering the word becomes easier.
Hence, if the attacker knows the presence of cache prefetchers in advance,
she can use the information to correlate the prefetched addresses with each word (\S\ref{sec:matching}).

As shown in Figure~\ref{fig:hunspell_confidence}, cache prefetching increases 
the average number of useful traces per word.
Including prefetched addresses increases the recovery rate
especially when there are very few useful traces (None and SQ).
Although the improvement is marginal in our experiment,
the attacker can potentially use the additional memory requests 
made by the cache prefetchers to extract more information from the victim.

\subsubsection{Memcached: Advantage of Fine-Grained Addresses}

To show the advantage of observing fine-grained addresses,
we simulated the controlled-channel attack on \memcached example.
We first obtained the entire memory trace from \memcached without simulating the cache.
We then masked the lower 12-bits of all addresses
assuming each page is 4~KB.
With this post-processing, we were able to simulate the memory trace
that the controlled-channel attacker will observe.
We also reconstruct the attacker's hash table such that each page-granularity address maps to multiple entries in the hash table.
If the attacker sees an address, she simply chooses the most common word among the
possible entries.

The simulated controlled-channel attack achieves only $29\%$ accuracy,
and the recovered document was uninterpretable as it only contained common words such as ``the'' and ``of''.
This shows that \scheme leverages fine-grained addresses by providing more side-channel information than coarse-grained addresses.

\section{Discussion}
\label{sec:discussion}

In this section, we discuss the limitations, generalization, implications, and mitigations of the \scheme{} attack.

\paragraph{Limitations.}
\label{sec:discussion:limitations}

\scheme{}  leaks only memory access patterns at LLC misses.
Thus, \scheme{} cannot observe repeated accesses to the same address within a short period.
For instance, the former RSA implementation of GnuPG~\cite{gnupg} is known to leak a private key through code addresses in the ElGamal algorithm~\cite{flush-reload}.
This type of attack relies on {\em data-dependent branches}, as the attacker detects different code paths executed inside the victim to infer the secret.
However, these vulnerabilities are difficult to exploit by \scheme{}, due to these code addresses being frequently executed and thus cached in the CPU.
Even cache priming techniques cannot efficiently evict the code addresses in time to help the attacker retrieve the secret with high accuracy but keep the performance impact low.

In general, \scheme{} is more suitable for leaking {\em data-dependent memory loads} over a large heap or array.
For instance, both the attacks on Hunspell and Memcached rely on the access patterns within a large hash table and/or linked-list objects.
If the victim program only has data-dependent memory access patterns within a small region, or if the memory access is not evenly distributed, the accuracy of \scheme{} is likely to worsen.
Besides, if the application only leaks a secret through {\em stores} that are dependent on the secret, \scheme{} may not observe the memory requests immediately.
The reason is that the CPU tends to delay {\em write-back} of dirty data until the cache lines are evicted, making the timing of the memory requests appearing on the memory bus unpredictable.
We leave the exploration of such scenario for future work.

\paragraph{Timing Information.}
Although not explored in this paper, an attacker may exploit the timing information to attack the victim.
The DRAM analyzer logs
a precise timestamp for each memory request based on counting its clock cycles.
Potentially, an attacker can measure the time difference
between two memory traces, to infer
the execution time of operation in the victim
as a way of timing attacks.
We leave the demonstration of these attacks for future work.

\paragraph{Traffic Analysis.}
Potentially, the memory bus traffic recorded by the DRAM analyzer can be used for traffic analysis
if the victim is vulnerable to this type of attacks.
For instance,
the attacker may analyze either the density
or the volume of requests on a specific address to infer the activity or secret of the application.
A complete mitigation of the attack should eliminate the timing information and has a constant traffic flow on the memory bus~\cite{InvisiMem}.

\paragraph{Multiple DIMMs or Multi-Socket.}
\label{sec:discussion:multi-dimm}
Our current attack does not explore the possibility of having multiple DIMMs or multiple CPU sockets (currently not supported by SGX).
However, potentially, the attacker can attach
multiple DIMM interposers, and then correlate the DRAM traces using timestamps or common patterns.

\paragraph{Memory Controllers.}

A memory controller 
arbitrates all transactions to main memory such that it maximizes the throughput while minimizing latencies.
One of the key features that may make \scheme{} more challenging is {\it transaction scheduling} where the {\it arbiter} reorders the transaction requests to maximize the performance.
In other words, the order of the memory transactions observed by the attacker may differ from the actual order of memory accesses.

We observe that the arbitration of the memory controller does not stop an enclave from leaking sensitive access patterns.
First, even if transactions are reordered, the critical addresses will still eventually appear on the memory bus.
Also, the memory controller only reorders transactions within a very small time window (e.g., tens of bus cycles), which is not enough to obfuscate the critical memory accesses that occur at least every hundreds of instructions.

\paragraph{Generalization.}
\label{sec:discussion:otherplatforms}

Intel SGX is not the only platform affected by \scheme{}.
Other existing platforms of hardware enclaves~\cite{komodo,sanctum,keystone,cryptoisland} also do not encrypt the addresses on the memory bus. Thus, these platforms are also vulnerable to \scheme{}
as long as the CPU stores encrypted data in external memory (e.g., DRAM).
The attacker can also use the same techniques such as cache squeezing 
to induce cache misses on other platforms.
For example, Komodo~\cite{komodo} allows the OS to affect the virtual address mapping, which enables the attacker to use \squeezing.
Keystone~\cite{keystone} measures the initial virtual address mapping for attestation, 
thus \squeezing cannot be applied. 
However, it provides cache partitioning which can reduce the effective cache size of the enclave.

\paragraph{Implications and Disclosure.}

Potentially, \scheme{} can be used in two scenarios:
(1) a malicious user attacking an end device to retrieve secret data from a local enclave;
(2) a malicious cloud provider or employee attacking a cloud machine to retrieve secret data from the tenants.
The existence of \scheme{} shows the importance of physical security to enclaves
just on par with software security. 
Ideally, in a secure cloud, one may want to separate the person who has physical access
to the machine from the person who has administrative privileges.
This may be achieved by a secure boot system
that prevents people who have physical access from overwriting system privileges.

We have disclosed the details of
this attack to Intel, who has acknowledged its validity.

\paragraph{Mitigations.}
\label{sec:discussion:mitigation}

There are several ways to mitigate \scheme{}, but they are generally expensive. 
Oblivious RAM (ORAM)~\cite{pathoram,zerotrace} can make the applications execute in an oblivious manner so that the attacker cannot infer secret data based on the memory access pattern.
The high performance overhead of ORAM makes it less attractive for applications 
that have strong performance requirements.
Alternatively, we can also encrypt the address bus as proposed by InvisiMem~\cite{InvisiMem} and ObfusMem~\cite{ObfusMem}.
However, adding such a feature to commodity DRAM would be very expensive; take the cost of techniques such as Hybrid Memory Cube (HMC)~\cite{hmc} for an example.
In-package memory such as high bandwidth memory (HBM) may relieve the needs for protection against untrusted DRAM~\cite{kommerling1999design}, but remains an expensive alternative for production.

\section{Conclusion}

In this paper, we introduced \scheme{}, which is a non-interference, fine-grained, stealthy physical side-channel attack on hardware enclaves based on snooping the address lines of the memory bus off-chip.
The key idea is to exploit OS privileges to induce cache misses with minimal performance overhead.
We also demystify the physical bus-based side channel by
reverse-engineering the internals of several hardware components.
We then develop an algorithm that can retrieve application secrets from memory bus traces. 
We demonstrated the attack on an actual SGX machine; the attack achieved similar accuracy with much lower overhead than previous attacks such as controlled-channel attacks.
We believe the attack technique is prevalent beyond Intel SGX and can apply to other secure processors or enclave platforms, which do not protect memory buses.

\section*{Acknowledgments}

We thank our shepherd, Daniel Genkin, and the anonymous reviewers for their insightful comments. We thank Krste Asanovi\'{c} and Martin Maas for sharing their ideas. Jeongseok Son from UC Berkeley also contributed to the early stage of the project.
We also thank SK Hynix, especially Dongha Jung, Taeksang Song, and Yongtak Song for providing the facility for DRAM signal analysis, collecting physical experiment data, and explaining the technical details of DRAM.
This work was supported in part by NSF grants CNS-1228839, CNS-1405641, CNS-1700512, NSF CISE Expeditions Award CCF-1730628, as well as gifts from the Sloan Foundation, Alibaba, Amazon Web Services, Ant Financial, ARM, Capital One, Ericsson, Facebook, Google, Intel, Microsoft, Scotiabank, Splunk, and VMware.

\bibliographystyle{unsrt}
\bibliography{reference.bib}

\label{LastPage}

\end{document}